\documentclass[preprint,onecolumn,aps,pre,eqsecnum,longbibliography,showpacs,showkeys,a4paper,floatfix]{revtex4-1}
\usepackage[latin9]{inputenc}
\usepackage{color}
\definecolor{note_fontcolor}{rgb}{0.80078125, 0.80078125, 0.80078125}
\usepackage[english]{babel}
\usepackage{amsmath}
\usepackage{amssymb}
\usepackage{graphicx}
\usepackage{esint}
\usepackage[unicode=true,
 bookmarks=true,bookmarksnumbered=true,bookmarksopen=true,bookmarksopenlevel=1,
 breaklinks=true,pdfborder={0 0 0},backref=false,colorlinks=true]
 {hyperref}
\hypersetup{
 pdfauthor={AINS}}

\makeatletter

\newenvironment{lyxgreyedout}
  {\textcolor{note_fontcolor}\bgroup\ignorespaces}
  {\ignorespacesafterend\egroup}	

\@ifundefined{textcolor}{}
{%
 \definecolor{BLACK}{gray}{0}
 \definecolor{WHITE}{gray}{1}
 \definecolor{RED}{rgb}{1,0,0}
 \definecolor{GREEN}{rgb}{0,1,0}
 \definecolor{BLUE}{rgb}{0,0,1}
 \definecolor{CYAN}{cmyk}{1,0,0,0}
 \definecolor{MAGENTA}{cmyk}{0,1,0,0}
 \definecolor{YELLOW}{cmyk}{0,0,1,0}
 }
\numberwithin{equation}{section}
\numberwithin{figure}{section}
\numberwithin{table}{section}

\usepackage{graphicx}

\DeclareGraphicsExtensions{.png,.pdf,.eps}
\graphicspath{{pictures/}}

\makeatother

\begin{document}

\title{Emergence of Quantum Mechanics from a Sub-Quantum Statistical Mechanics}

\author{Gerhard \surname{Grössing}}

\email[E-mail: ]{ains@chello.at}

\homepage[Visit: ]{http://www.nonlinearstudies.at/}

\affiliation{Austrian Institute for Nonlinear Studies, Akademiehof\\
 Friedrichstr.~10, 1010 Vienna, Austria\vspace*{3cm}
}

\begin{abstract}
A research program within the scope of theories on \textquotedblleft{}Emergent
Quantum Mechanics\textquotedblright{} is presented, which has gained
some momentum in recent years. Via the modeling of a quantum system
as a non-equilibrium steady-state maintained by a permanent throughput
of energy from the zero-point vacuum, the quantum is considered as
an \emph{emergent system}. We implement a specific \textquotedblleft{}bouncer-walker\textquotedblright{}
model in the context of an assumed sub-quantum statistical physics,
in analogy to the results of experiments by Couder\textquoteright{}s
group on a classical wave-particle duality. We can thus give an explanation
of various quantum mechanical features and results on the basis of
a \textquotedblleft{}21\textsuperscript{st} century classical physics\textquotedblright{},
such as the appearance of Planck\textquoteright{}s constant, the Schrödinger
equation, etc. An essential result is given by the proof that averaged
particle trajectories\textquoteright{} behaviors correspond to a specific
type of anomalous diffusion termed \textquotedblleft{}ballistic\textquotedblright{}
diffusion on a sub-quantum level.

It is further demonstrated both analytically and with the aid of computer
simulations that our model provides explanations for various quantum
effects such as double-slit or \emph{n}-slit interference. We show
the averaged trajectories emerging from our model to be identical
to Bohmian trajectories, albeit without the need to invoke complex
wave functions or any other quantum mechanical tool. Finally, the
model provides new insights into the origins of entanglement, and,
in particular, into the phenomenon of a ``systemic'' nonlocality.%
\begin{lyxgreyedout}
\global\long\def\VEC#1{\mathbf{#1}}

\global\long\def\d{\,\mathrm{d}}

\global\long\def\e{{\rm e}}

\global\long\def\i{{\rm i}}

\global\long\def\meant#1{\left<#1\right>}

\global\long\def\meanx#1{\overline{#1}}

\global\long\def\mpbracket{\ensuremath{\genfrac{}{}{0pt}{1}{-}{\scriptstyle (\kern-1pt +\kern-1pt )}}}

\global\long\def\pmbracket{\ensuremath{\genfrac{}{}{0pt}{1}{+}{\scriptstyle (\kern-1pt -\kern-1pt )}}}

\global\long\def\p{\partial}
\end{lyxgreyedout}

\end{abstract}


\maketitle

\section{Introduction}

After many attempts to frame wave-particle duality on a single ``basic''
explanatory level, like on pure formalism (Heisenberg,\ldots{}),
pure wave mechanics (de\,Broglie,\ldots{}), or pure particle physics
(Feynman,\ldots{}), for example, it is instructive to ask the following
question: is perhaps the quantum a more complex \emph{dynamical} phenomenon?
One motivation for raising this question comes from the fact that
we are presently witnessing a historical change with respect to wave-particle
duality in double-slit interference. With regard to the latter, the
standard attitude during practically the whole of the 20th century
can be characterized by Richard Feynman's famous claim that it is
``... a phenomenon which is impossible, absolutely impossible to
explain in any classical way.'' In the present, however, we are facing
the beautiful experiments by Yves Couder's group in Paris which do
show wave-particle duality in a completely classical system, i.e.,
with the aid of bouncers/walkers on an oscillating fluid. (For a first
impression, see the video with Yves Couder \cite{Couder.2011yves}.)

In fact, with our approach we have in a series of papers obtained
from a ``classical'' dynamical model several essential elements
of quantum theory \cite{Groessing.2008vacuum,Groessing.2009origin,Groessing.2010emergence,Groessing.2011dice,Groessing.2011explan,Groessing.2012doubleslit,Grossing.2012quantum}.
They derive from the assumption that a ``particle'' of energy $E=\hbar\omega$
is actually an oscillator of angular frequency $\omega$ phase-locked
with the zero-point oscillations of the surrounding environment, the
latter of which containing both regular and fluctuating components
and being constrained by the boundary conditions of the experimental
setup via the buildup and maintenance of standing waves. The ``particle''
in this approach is an off-equilibrium steady-state maintained by
the throughput of zero-point energy from its ``vacuum'' surroundings.
In other words, in our model a quantum \emph{emerges} from the synchronized
dynamical coupling between a bouncer and its wave-like environment.
This is in close analogy to the bouncing/walking droplets in the experiments
of Couder's group \cite{Couder.2005,Couder.2006single-particle,Couder.2010walking,Couder.2012probabilities},
which in many respects can serve as a classical prototype guiding
our intuition. Our research program thus pertains to the scope of
theories on ``Emergent Quantum Mechanics''. (For the proceedings
of a first international conference exclusively devoted to this topic,
see Grössing (2012) \cite{Groessing.2012emerqum11-book}. For their
original models, see in particular the papers by Adler, Elze, Ord,
Grössing\emph{~et\,al}., Cetto\emph{~et\,al}., Hiley, de\,Gosson,
Bacciagaluppi, Budiyono, Schuch, Faber, Hofer, 't\,Hooft, Khrennikov\emph{~et\,al}.,
Wetterich, and Burinskii.) We have recently applied our model to the
case of interference at a double slit \cite{Groessing.2012doubleslit},
thereby obtaining the exact quantum mechanical probability distributions
on a screen behind the double slit, the average particle trajectories
(which because of the averaging are shown to be identical to the Bohmian
ones), and the involved probability density currents.

Moreover, already some decades ago, Aharonov\emph{~et\,al}.\ \cite{Aharonov.1969modular,Aharonov.1970deterministic}
investigated the problem of double-slit interference ``from a single
particle perspective''. Their explanation is based on what they call
\emph{dynamical nonlocality}, and they therewith answer the question
of how a particle going through one slit can ``know'' about the
state of the other slit (i.e., being open or closed, for example).
This type of ``dynamical'' nonlocality may lead to (causality preserving)
changes in probability distributions and is thus distinguished from
the \emph{kinematic nonlocality} implicit in many quantum correlations,
which, however, does not cause any changes in probability distributions.
Chapter 3 of the present review extends the applicability also of
our model to a such a ``dynamical'' scenario, which in our terminology
rather relates to a ``systemic'' nonlocality, as shall be explicated
below.

This paper is organized as follows. In Chapter 2, some main results
of our sub-quantum approach to quantum mechanics are presented. We
begin by constructing from classical physics a model that explains
the quantum mechanical dispersion of a Gaussian wave packet. As a
consequence, we obtain the total velocity field for the current emerging
from a Gaussian slit, which can easily be extended to a two-slit (and
in fact to any \emph{n}-slit) system, thereby also providing an explanation
of the famous interference effects. With the thus obtained velocity
field, one can by integration also obtain an action function that
generalizes the ordinary classical one to the case that is relevant
for the reproduction of the quantum results. Effectively, this is
achieved by the appearance of an additional kinetic energy term, which
we associate with fluxes of ``heat'' in the vacuum, i.e., sub-quantum
currents based on the existence of the zero-point energy field. Finally,
it is shown how the Schrödinger equation can be derived in a straightforward
manner from the said generalized action function, or the corresponding
Lagrangian, respectively. In Chapter 3, then, we deal with the above-mentioned
problem of interference from the single particle perspective, obtaining
results which are in agreement with those presented by Aharonov\emph{~et\,al}.\ \cite{Aharonov.1969modular,Aharonov.1970deterministic},
but differing in interpretation. Finally, in Chapter 4 computer simulations
are presented to show some examples of how our sub-quantum approach
reproduces results of ordinary quantum mechanics, but eventually may
go beyond these.

\section{The Quantum as an Emergent System: A Sub-Quantum Approach to Quantum
Mechanics\label{sec:A-short-Introduction}}

We transfer an insight from the Couder experiments into our modeling
of quantum systems and assume that the waves are a space-filling phenomenon
involving the whole experimental setup. Thus, one can imagine a partial
decoupling of the physics of waves and particles in that the latter
still may be ``guided'' through said ``landscape'', but the former
may influence other regions of the ``landscape'' by providing specific
phase information independently of the propagation of the particle.
This is why a remote change in the experimental setup, when mediated
to the particle via de- and/or re-construction of standing waves,
can amount to a nonlocal effect on a particle via the thus modified
guiding landscape.

In earlier work \cite{Groessing.2010emergence} we presented a model
for the classical explanation of the quantum mechanical dispersion
of a free Gaussian wave packet. In accordance with the classical model,
we now relate it more directly to a ``double solution'' analogy
gleaned from Couder and Fort \cite{Couder.2012probabilities}. These
authors used the double solution ansatz to describe the behaviors
of their ``bouncer''- (or ``walker''-) droplets: on an individual
level, one observes particles surrounded by circular waves they emit
through the phase-coupling with an oscillating bath, which provides,
on a statistical level, the emergent outcome in close analogy to quantum
mechanical behavior (like, e.g., diffraction or double-slit interference).
Originally, the expression of a ``double solution'' refers to an
early idea of de\,Broglie \cite{DeBroglie.1960book} to model quantum
behavior by a two-fold process, i.e., by the movement of a hypothetical
point-like ``singularity solution'' of the Schrödinger equation,
and by the evolution of the usual wavefunction that would provide
the empirically confirmed statistical predictions. Now, it is interesting
to observe that one can construct various forms of classical analogies
to the quantum mechanical Gaussian dispersion \cite{Holland.1993},
and one of them even can be related to the double solution idea. 

To establish correlations on a statistical level between individual
uncorrelated particle positions $x$ and momenta $p$, respectively,
one considers a solution of the free Liouville equation providing
a certain phase-space distribution $f\left(x,p,t\right)$. This distribution
shows the emergence of correlations between $x$ and $p$ from an
initially uncorrelated product function of non-spreading (``classical'')
Gaussian position distributions as well as momentum distributions.
The motivation for their introduction comes exactly from what one
observes in the Couder experiments. In an idealized scenario, we assume
that at each point $x$ an unbiased emission of momentum fluctuations
$\pi_{0}$ in all possible directions takes place, thus mimicking
(in a two-dimensional scenario) the circular waves emitted from the
``particle as bouncer''. If we compare the typical frequency of
the bouncers in the Couder experiments (i.e., roughly $10^{2}\mathrm{\, Hz}$)
with that of an electron, for example (i.e., roughly $10^{20}\,\mathrm{Hz}$),
we see that a ``continuum ansatz'' is very pragmatical, particularly
if we are interested in statistical averages over a long series of
experimental runs. 

Thus, one can construct said phase-space distribution, with $\sigma_{0}$
being the initial $x$--space standard deviation, i.e., $\sigma_{0}=\sigma(t=0)$,
and $\pi_{0}:=mu_{0}$ the momentum standard deviation, such that
\begin{equation}
f\left(x,p,t\right)=\frac{1}{2\pi\sigma_{0}mu_{0}}\exp\left\{ -\frac{\left(x-pt/m\right)^{2}}{2\sigma_{0}^{2}}\right\} \exp\left\{ -\frac{p^{2}}{2m^{2}u_{0}^{2}}\right\} .\label{eq:traj.2.6}
\end{equation}

\noindent Now, the above-mentioned correlations between $x$ and $p$
emerge when one considers the probability density in $x$--space.
Integration over $p$ provides
\begin{equation}
P\left(x,t\right)=\int f\d p=\frac{1}{\sqrt{2\pi}\sigma}\exp\left\{ -\frac{x^{2}}{2\sigma^{2}}\right\} ,\label{eq:traj.2.7}
\end{equation}
with the standard deviation at time $t$ given by
\begin{equation}
\sigma^{2}=\sigma_{0}^{2}+u_{0}^{2}\, t^{2}.\label{eq:traj.2.8}
\end{equation}

In other words, the distribution \eqref{eq:traj.2.7} with property
\eqref{eq:traj.2.8} describes a spreading Gaussian which is obtained
from a continuous set of classical, i.e., non-spreading, Gaussian
position distributions of particles whose associated momentum fluctuations
also have non-spreading Gaussian distributions. One thus obtains the
exact quantum mechanical dispersion formula for a Gaussian, as we
have obtained also previously from a different variant of our classical
ansatz. For confirmation with respect to the latter (diffusion-based)
model \cite{Groessing.2010emergence,Groessing.2011dice}, we consider
with \eqref{eq:traj.2.7} the usual definition of the ``osmotic''
velocity field, which in this case yields $u=-D\frac{\nabla P}{P}=\frac{xD}{\sigma^{2}}$.
One then obtains, with bars denoting averages over fluctuations and
positions, i.e., with $\meanx{x^{2}}=\int Px^{2}dx=\sigma^{2}$,
\begin{equation}
\meanx{u^{2}}=D^{2}\meanx{\left(\frac{\nabla P}{P}\right)^{2}}=D^{2}\int P\left(\frac{\nabla P}{P}\right)^{2}\d x=\frac{D^{2}}{\sigma^{2}},\;\;\textrm{and thus also}\;\; u_{0}=\frac{D}{\sigma_{0}},\label{eq:traj.2.9}
\end{equation}

\noindent so that one can rewrite Eq.~\eqref{eq:traj.2.8} in the
more familiar form
\begin{equation}
\sigma^{2}=\sigma_{0}^{2}\left(1+\frac{D^{2}t^{2}}{\sigma_{0}^{4}}\right).\label{eq:traj.2.10}
\end{equation}

\noindent Note also that by using the Einstein relation $D=\hbar/2m=h/4\pi m$
the norm in \eqref{eq:traj.2.6} thus becomes the invariant expression
(reflecting the ``exact uncertainty relation'' \cite{Hall.2002schrodinger})
\begin{equation}
\frac{1}{2\pi\sigma_{0}mu_{0}}=\frac{1}{2\pi mD}=\frac{2}{h}.\label{eq:traj.2.11}
\end{equation}

Following from \eqref{eq:traj.2.10}, in Grössing\emph{~et\,al}.\ \cite{Groessing.2010emergence,Groessing.2012doubleslit}
we obtained for ``smoothed-out'' trajectories (i.e., averaged over
a very large number of Brownian motions) a sum over a deterministic
and a fluctuations term, respectively, for the motion in the $x$--direction
\begin{equation}
x_{{\rm tot}}(t)=vt+x(t)=vt+x(0)\frac{\sigma}{\sigma_{0}}=vt+x(0)\sqrt{1+\frac{D^{2}t^{2}}{\sigma_{0}^{4}}}.\label{eq:2.22}
\end{equation}
Thus one classically obtains with $x\left(t\right)=x_{{\rm tot}}(t)-vt$
the \emph{average velocity field} of a Gaussian wave packet as 
\begin{equation}
v_{{\rm tot}}(t)=v(t)+\frac{\d x(t)}{{\rm \d t}}=v(t)+\left[x(t)\right]\frac{u_{0}^{2}t}{\sigma^{2}}.\label{eq:2.23}
\end{equation}
For the particular choice of $x\left(t\right)=\sigma$, then, and
with $\frac{\sigma}{t}\rightarrow u_{0}$ for $t\rightarrow\infty$,
one obtains for large $t$ that $v_{{\rm tot}}(t)=v(t)+u_{0}$.

Note that Eqs.~\eqref{eq:2.22} and \eqref{eq:2.23} are derived
solely from statistical physics. Still, they are in full accordance
with quantum theory, and in particular with Bohmian trajectories \cite{Holland.1993}.
Note also that one can rewrite Eq.~\eqref{eq:traj.2.10} such that
it appears like a linear-in-time formula for Brownian motion, 
\begin{equation}
\meanx{x^{2}}=\meanx{x^{2}(0)}+D_{\mathrm{t}}\, t,\label{eq:2.24}
\end{equation}
where a time dependent diffusivity 
\begin{equation}
D_{\mathrm{t}}=u_{0}^{2}\, t=\frac{\hbar^{2}}{4m^{2}\sigma_{0}^{2}}\, t\label{eq:2.25}
\end{equation}
characterizes Eq.~\eqref{eq:2.24} as \textit{ballistic diffusion}.
This makes it possible to simulate the dispersion of a Gaussian wave
packet on a computer by simply employing coupled map lattices for
classical diffusion, with the diffusivity given by Eq.~\eqref{eq:2.25}.
(For detailed discussions, see Grössing\emph{~et\,al}.\ (2010)
\cite{Groessing.2010emergence} and Grössing\emph{~et\,al}.\ (2011)
\cite{Groessing.2011dice} and the last Chapter of this paper.)

Moreover, one can easily extend this scheme to more than one slit,
like, for example, to explain interference effects at the double slit
\cite{Groessing.2012doubleslit,Grossing.2012quantum}. For this, we
chose similar initial situations as in Holland \cite{Holland.1993},
i.e., electrons (represented by plane waves in the forward $y$--direction)
from a source passing through ``soft-edged'' slits $1$ and $2$
in a barrier (located along the $x$--axis) and recorded at a screen.
In our model, we therefore note two Gaussians representing the totality
of the effectively ``heated-up'' path excitation field, one for
slit $1$ and one for slit $2$, whose centers have the distances
$+X$ and $-X$ from the plane spanned by the source and the center
of the barrier along the $y$--axis, respectively. 

With the total amplitude $R$ of two coherent waves with (suitably
normalized) amplitudes $R_{i}=\sqrt{P_{i}}$, and the local phases
$\varphi_{i}$, $i=1$ or $2$, one has, according to classical textbook
wisdom, the \emph{averaged total intensity}
\begin{equation}
P_{{\rm tot}}:=R^{2}=R_{1}^{2}+R_{2}^{2}+2R_{1}R_{2}\cos\varphi_{12}=P_{1}+P_{2}+2\sqrt{P_{1}P_{2}}\cos\varphi_{12},\label{eq:3.0a-1-1}
\end{equation}
where $\varphi_{12}$ is the relative phase $\varphi_{12}=\varphi_{1}-\varphi_{2}=\left(\VEC k_{1}-\VEC k_{2}\right)\cdot\VEC r$.
Note that $\varphi_{12}$ enters Eq.~\eqref{eq:3.0a-1-1} only via
the cosine function, such that, e.g., even if the total wave numbers
(and thus also the total momenta) $\VEC k_{i}$ were of vastly different
size, the cosine effectively makes Eq.~\eqref{eq:3.0a-1-1} independent
of said sizes, but dependent only on an angle modulo $2\pi$. This
will turn out as essential for our discussion further below.

The $x$--components of the centroids' motions from the two alternative
slits $1$ and $2$, respectively, are given by the ``particle''
velocity components 
\begin{equation}
v_{x}=\pm\frac{\hbar}{m}\, k_{x},\label{eq:3.1-1-1}
\end{equation}
respectively, such that the relative group velocity of the Gaussians
spreading into each other is given by $\Delta v_{x}=2v_{x}$. However,
in order to calculate the phase difference $\varphi_{12}$ descriptive
of the interference term of the intensity distribution \eqref{eq:3.0a-1-1},
one must take into account the total momenta involved, i.e., one must
also include the wave packet dispersion as described above. Thus,
one obtains with the displacement $\pm x\left(t\right)=\mp\left(X+v_{x}t\right)$
in Eq.~\eqref{eq:2.23} the total relative velocity of the two Gaussians
as 
\begin{equation}
\Delta v_{{\rm tot},x}=2\left[v_{x}-(X+v_{x}t)\frac{u_{0}^{2}t}{\sigma^{2}}\right].\label{eq:3.1a-1-1}
\end{equation}
Therefore, the total phase difference between the two possible paths
1 and 2 (i.e., through either slit) becomes 
\begin{equation}
\varphi_{12}=\frac{1}{\hbar}(m\Delta v_{{\rm tot},x}\, x)=2mv_{x}\frac{x}{\hbar}-(X+v_{x}t)x\frac{1}{D}\frac{u_{0}^{2}t}{\sigma^{2}}.\label{eq:dyn.2.19}
\end{equation}

In earlier papers, Grössing~\emph{(et\,al}.) \cite{Groessing.2008vacuum,Groessing.2009origin,Groessing.2010emergence},
we have shown that, apart from the ordinary particle current $\VEC J(\VEC x,t)=P(\VEC x,t)\VEC v$,
we are now dealing with two additional, yet opposing, currents $\VEC J_{u}=P(\VEC x,t)\VEC u$,
which are on average orthogonal to $\VEC J$, and which are the emergent
outcome from the presence of numerous momentum fluctuations, and the
corresponding velocities, 
\begin{equation}
\VEC u_{\pm}=\mp\frac{\hbar}{2m}\frac{\nabla P}{P}.\label{eq:2.7}
\end{equation}
We denote with $\VEC u_{+}$ and $\VEC u_{-}$, respectively, the
two opposing tendencies of the diffusion process. Moreover, one can
also take the averages over fluctuations and positions to obtain a
``smoothed-out'' \textit{average velocity field} \cite{Groessing.2008vacuum,Groessing.2009origin,Groessing.2010emergence,Groessing.2011dice},
i.e., 
\begin{equation}
\meanx{\VEC u(\VEC x,t)}=\int P\VEC u(\VEC x,t)\d^{n}x.\label{eq:2.10}
\end{equation}

In effect, the combined presence of the velocity fields $\VEC u$
and $\VEC v$ can be denoted as a \textit{path excitation field}:
via diffusion, the bouncer in its interaction with already existing
wave-like excitations of the environment creates an ``agitated'',
or ``heated-up'', thermal landscape, which can also be pictured
by interacting wave configurations all along between source and detector
of an experimental setup. Recall that our prototype of a ``walking
bouncer'', i.e., from the experiments of Couder's group, is always
driven by its interactions with a superposition of waves emitted at
the points it visited in the past. Couder\emph{~et\,al}.\ denote
this superposition of in-phase waves the ``path memory'' of the
bouncer \cite{Fort.2010path-memory}. This implies, however, that
the bouncers at the points visited in ``the present'' necessarily
create new wave configurations which will form the basis of a ``path
memory'' in the future. In other words, the wave configurations of
the past determine the bouncer's path in the present, whereas its
bounces in the present co-determine the wave configurations, and thus
the probability density distribution, at any of the possible locations
it will visit in the future. Therefore, by calling the latter configurations
a \textit{path excitation field}, we point to our model's physical
meaning of the probability density distribution: its time evolution
is to be understood as the totality of all sub-quantum currents, which
may also be described as a ``heated-up'' thermal field. 

For illustration, let us now consider a single, classical ``particle''
(``bouncer'') following the propagation of a set of waves of equal
amplitude $R_{i}$, each representing one of $i$ possible alternatives
according to our principle of path excitation, and focus on the specific
role of the velocity fields. To describe the required details, each
path $i$ be occupied by a Gaussian wave packet with a ``forward''
momentum $\VEC p_{i}=\hbar\VEC k_{i}=m\VEC v_{i}$. Moreover, due
to the stochastic process of path excitation, the latter has to be
represented also by a large number $N$ of consecutive Brownian shifts
as in Eq. \eqref{eq:2.7}, which on average, for stretches of free
particle propagation $\VEC v_{i}$, are orthogonal to $\VEC v_{i}$.
Defining an average total velocity (with indices $i=1$ or $2$ referring
to the two slits, and with $+$ and $-$ referring to the right and
the left from the average direction of $\VEC v_{i}$, respectively)
\begin{equation}
\meanx{\VEC v}_{{\rm tot},i}:=\meanx{\VEC v}_{i}+\meanx{\VEC u}_{i+}+\meanx{\VEC u}_{i-,}\label{eq:3.8}
\end{equation}
with the bars here (and further on, if not declared otherwise) denoting
averages only over the spatial directions, we consider two Gaussian
probability density distributions, $P_{1}=R_{1}^{2}$ and $P_{2}=R_{2}^{2}$,
respectively. Generally, with the total amplitude $R$ of two coherent
waves with (suitably normalized) amplitudes $R_{i}=\sqrt{P_{i}}$,
and the local phases $\varphi_{i}$, $i=1$ or $2$, one has as usual
that
\begin{equation}
R=R_{1}\cos\left(\omega t+\varphi_{1}\right)+R_{2}\cos\left(\omega t+\varphi_{2}\right),\label{eq:11a}
\end{equation}
which, when squared and averaged, provides the famous formula for
the intensity of the interference pattern \eqref{eq:3.0a-1-1}. Introducing
an arbitrarily chosen unit vector $\mathbf{\hat{n}}$, one may also
define $\cos\left(\omega t+\varphi_{i}\left(\mathbf{x}\right)\right)$=
$\mathbf{\hat{n}\cdot\mathbf{\hat{k_{i}}\left(\mathbf{x}\mathrm{,t}\right)}},$
such that along with the system's evolution, the emergent outcome
of the time evolution of $R\left(\mathbf{x}\mathrm{,t}\right)$ as
a characteristic of our path excitation field can analogously be assumed
as 
\begin{equation}
R\left(\mathbf{x}\mathrm{,t}\right)=\hat{\mathbf{n}}\cdot\left(R_{1}\mathbf{\hat{k}}_{1}\left(\mathbf{x}\mathrm{,t}\right)+R_{2}\mathbf{\hat{k}}_{2}\left(\mathbf{x}\mathrm{,t}\right)\right).\label{eq:11b}
\end{equation}
Thus, with regard to the total amplitude $R_{\mathrm{tot}}$ in the
double-slit case, one obtains with the appropriate normalization that
\begin{equation}
R_{\mathrm{tot}}^{2}=\left(R_{1}\meanx{\hat{\VEC v}}_{{\rm tot},1}+R_{2}\meanx{\hat{\VEC v}}_{{\rm tot},2}\right)^{2}.
\end{equation}
With $P_{\mathrm{tot}}=R_{\mathrm{tot}}^{2}$, one can calculate the
time development of the path excitation field, i.e., the spreading
out of the total probability density in the form of the total average
current $\meanx{\VEC J}_{{\rm tot}}=P_{\mathrm{tot}}\meanx{\VEC v}_{\mathrm{tot}}$.
After a few calculational steps this provides, similarly to Grössing\emph{~et\,al}.\ (2012)\cite{Groessing.2012doubleslit}
(but now with slightly different labellings which apply more generally),
with the assumed orthogonality $\meanx{\VEC v}_{i}\cdot\meanx{\VEC u}_{i}=0$
for free particle propagation ($i=1\textrm{ or }2$) that 
\begin{equation}
\meanx{\VEC J}_{{\rm tot}}=P_{1}\meanx{\VEC v}_{1}+P_{2}\meanx{\VEC v}_{2}+\sqrt{P_{1}P_{2}}\left(\meanx{\VEC v}_{1}+\meanx{\VEC v}_{2}\right)\cos\varphi_{12}+\sqrt{P_{1}P_{2}}\left(\meanx{\VEC u}_{1}-\meanx{\VEC u}_{2}\right)\sin\varphi_{12}.\label{eq:3.12}
\end{equation}
An alternative, more detailed account of Eq.~\eqref{eq:3.12} and
its extension to \emph{n} slits is in preparation {[}Fussy~\emph{et\,al}.\ (2013){]}.
Note that Eq.~\eqref{eq:3.12}, upon the identification of $\meanx{\VEC u}_{i}=-\frac{\hbar}{m}\frac{\nabla R_{i}}{R_{i}}$
from Eq.~\eqref{eq:2.7} and with $P_{i}=R_{i}^{2}$, turns out to
be in perfect agreement with a comparable ``Bohmian'' derivation
\cite{Holland.1993,Sanz.2008trajectory}. In fact, with $\meanx{\VEC v}_{i}=$$\nabla S_{i}$/m,
one can rewrite \eqref{eq:3.12} as
\begin{align}
\meanx{\VEC J}_{{\rm tot}}= & \; R_{1}^{2}\frac{\nabla S_{1}}{m}+R_{2}^{2}\frac{\nabla S_{2}}{m}\label{eq:dyn.3.12.1}\\
 & +R_{1}R_{2}\left(\frac{\nabla S_{1}}{m}+\frac{\nabla S_{2}}{m}\right)\cos\varphi_{12}+\frac{\hbar}{m}\left(R_{1}\nabla R_{2}-R_{2}\nabla R_{1}\right)\sin\varphi_{12}.\nonumber 
\end{align}
The formula for the averaged particle trajectories, then, simply results
from 
\begin{equation}
\meanx{\VEC v}_{{\rm tot}}=\frac{\meanx{\VEC J}_{{\rm tot}}}{P_{\textrm{tot}}}.\label{eq:3.13}
\end{equation}

Although we have obtained the usual quantum mechanical results, we
have so far not used the quantum mechanical formalism in any way.
However, upon employment of the Madelung transformation for each path
$j$ ($j=1$ or $2$), 
\begin{equation}
\Psi_{j}=R\e^{\i S/\hbar},\label{eq:3.14}
\end{equation}
 and thus $P_{j}=R_{j}^{2}=|\Psi_{j}|^{2}=\Psi_{j}^{*}\Psi_{j}$,
with the definitions \eqref{eq:2.7} and $\overline{v_{j}}:=\nabla S_{j}/m$,
$\varphi_{12}=(S_{1}-S_{2})/\hbar$, and recalling the usual trigonometric
identities such as $\cos\varphi=\frac{1}{2}\left(\e^{\i\varphi}+\e^{-\i\varphi}\right)$,
etc., one can rewrite the total average current \eqref{eq:3.12} immediately
as
\begin{align}
{\displaystyle \meanx{\VEC J}_{{\rm tot}}} & =P_{{\rm tot}}\meanx{\VEC v}_{{\rm tot}}\nonumber \\
 & ={\displaystyle (\Psi_{1}+\Psi_{2})^{*}(\Psi_{1}+\Psi_{2})\frac{1}{2}\left[\frac{1}{m}\left(-\i\hbar\frac{\nabla(\Psi_{1}+\Psi_{2})}{(\Psi_{1}+\Psi_{2})}\right)+\frac{1}{m}\left(\i\hbar\frac{\nabla(\Psi_{1}+\Psi_{2})^{*}}{(\Psi_{1}+\Psi_{2})^{*}}\right)\right]}\nonumber \\
 & ={\displaystyle -\frac{\i\hbar}{2m}\left[\Psi^{*}\nabla\Psi-\Psi\nabla\Psi^{*}\right]={\displaystyle \frac{1}{m}{\rm Re}\left\{ \Psi^{*}(-\i\hbar\nabla)\Psi\right\} ,}}\label{eq:3.18}
\end{align}
where $P_{{\rm tot}}=|\Psi_{1}+\Psi_{2}|^{2}=:|\Psi|^{2}$. The last
two expressions of \eqref{eq:3.18} are the exact well-known formulations
of the quantum mechanical probability current, here obtained without
any quantum mechanics, but just by a re-formulation of \eqref{eq:3.12}.
In fact, it is a simple exercise to insert the wave functions \eqref{eq:3.14}
into \eqref{eq:3.18} to re-obtain \eqref{eq:3.12}.

Moreover, having thus obtained a ``bridge'' to the quantum mechanical
formalism, one can now show how the Schrödinger equation not only
complies with our classical ansatz, but actually can be derived from
it. Two different ways of such a corresponding derivation have already
been published by the present author (i.e., one on more general grounds
\cite{Groessing.2004hamilton}, the other within a more concrete model
based on nonequilibrium thermodynamics\cite{Groessing.2008vacuum}).
Here, however, we just start with a result from the construction of
the classical Gaussian with its dispersion mimicking the quantum mechanical
one, i.e., more concretely, with the resulting velocity field $v_{{\rm tot}}(t)$
from Eq.~\eqref{eq:2.23}. Integrating the latter, and with $\xi\left(t\right):=x-vt$
describing the location of a particle in a Gaussian probability density
distribution $P\left(x,t\right)=\frac{1}{\sqrt{2\pi}\sigma}\exp\left\{ -\frac{\left(x-vt\right)^{2}}{2\sigma^{2}}\right\} $,
we find an expression for the action function as
\begin{align}
S & =\int mv_{{\rm tot}}(t)\d x-\int E\d t\nonumber \\
 & =mvx+\frac{mu_{0}^{2}}{2}\left[\frac{\xi(t)}{\sigma(t)}\right]^{2}t-Et=mvx+\frac{mu_{0}^{2}}{2}\left[\frac{\xi(0)}{\sigma_{0}}\right]^{2}t-Et,\label{eq:2.6}
\end{align}
with $E$ being the system's total energy. Note that because it generally
holds that $\frac{\xi(t)}{\sigma(t)}=\frac{\xi(0)}{\sigma_{0}}$,
the physics contained in \eqref{eq:2.6} for the free particle case
is essentially determined already by the initial conditions, i.e.,
where in the Gaussian the particle is initially located. Note also
that the new term, as opposed to ``ordinary'' classical physics,
is given by a kinetic energy, which represents a thermal fluctuation
field with the kinetic temperature $kT=mu_{0}^{2}\left[\frac{\xi(t)}{\sigma(t)}\right]^{2}$
of what we have termed the \emph{path excitation field}. Now, as with
the above-mentioned Gaussian one has $u\left(x,t\right)=-D\frac{\nabla P}{P}=\frac{\xi(t)D}{\sigma^{2}}=\frac{\xi(t)\sigma_{0}}{\sigma^{2}}u_{0}$,
one can rewrite the new kinetic energy term in Eq.~\eqref{eq:2.6}
as
\begin{equation}
\frac{mu_{0}^{2}}{2}\left[\frac{\xi(t)}{\sigma(t)}\right]^{2}t=\frac{mu^{2}}{2}\left[\frac{\sigma(t)}{\sigma_{0}}\right]^{2}t=:\frac{m\tilde{u}^{2}}{2}t,
\end{equation}
where $\tilde{u}=u\frac{\sigma}{\sigma_{0}}$. (This is therefore
in complete accordance with the starting assumption in Grössing (2008)
\cite{Groessing.2008vacuum}.) Moreover, upon averaging over fluctuations
and space, one obtains that 
\begin{equation}
\int P\frac{mu_{0}^{2}}{2}\left[\frac{\xi(t)}{\sigma(t)}\right]dx=\frac{mu_{0}^{2}}{2}\equiv\frac{\hbar\omega}{2},
\end{equation}
where the latter equation describes the identity of the averaged kinetic
energy term with the zero-point energy as shown in Grössing (2009)
\cite{Groessing.2009origin}.

In more general terms, i.e., independently from the particular choice
of $P$ as a Gaussian distribution density, with our expression for
the momentum fluctuation $\delta\mathbf{p}=m\mathbf{u}$ as 
\begin{equation}
\delta\VEC p(\VEC x,t)=\nabla(\delta S(\VEC x,t))=-\frac{\hbar}{2}\frac{\nabla P(\VEC x,t)}{P(\VEC x,t)}\;.\label{eq:2.7-1}
\end{equation}
we can write our additional kinetic energy term for one particle as
\begin{equation}
\delta E_{{\rm kin}}=\frac{1}{2m}\nabla(\delta S\cdot\nabla(\delta S)=\frac{1}{2m}\left(\frac{\hbar}{2}\frac{\nabla P}{P}\right)^{2}\;.\label{eq:2.8}
\end{equation}
Thus, writing down a classical action integral for $n$ particles,
and including this new term for each of them, yields (with Lagrangian
$L$ and external potential $V$) 
\begin{align}
A & =\int L\d^{m}x\d t\nonumber \\
 & =\int P\left[\frac{\partial S}{\partial t}+\sum_{i=1}^{n}\frac{1}{2m_{i}}\nabla_{i}S\cdot\nabla_{i}S+\sum_{i=1}^{n}\frac{1}{2m_{i}}\left(\frac{\hbar}{2}\frac{\nabla_{i}P}{P}\right)^{2}+V\right]\d^{m}x\d t,\label{eq:2.9}
\end{align}
where $P=P(\VEC x_{1},\VEC x_{2},\ldots,\VEC x_{n},t)$. Using again
the Madelung transformation \eqref{eq:3.14} where $R=\sqrt{P}$,
one has 
\begin{equation}
\overline{\left|\frac{\nabla_{i}\psi}{\psi}\right|^{2}}:=\int\d^{m}x\d t\left|\frac{\nabla_{i}\psi}{\psi}\right|^{2}=\overline{\left(\frac{1}{2}\frac{\nabla_{i}P}{P}\right)^{2}}+\overline{\left(\frac{\nabla_{i}S}{\hbar}\right)^{2}}\;,\label{eq:2.11-1}
\end{equation}
and one can rewrite \eqref{eq:2.9} as 
\begin{equation}
A=\int L\d t=\int\d^{m}x\d t\left[|\psi|^{2}\left(\frac{\partial S}{\partial t}+V\right)+\sum_{i=1}^{n}\frac{\hbar^{2}}{2m_{i}}|\nabla_{i}\psi|^{2}\right]\;.\label{eq:2.12-1}
\end{equation}
Thus, with the identity $|\psi|^{2}\frac{\partial S}{\partial t}=-\frac{\i\hbar}{2}(\psi^{*}\dot{\psi}-\dot{\psi}^{*}\psi)$,
one obtains the familiar Lagrange density 
\begin{equation}
L=-\frac{\i\hbar}{2}(\psi^{*}\dot{\psi}-\dot{\psi}^{*}\psi)+\sum_{i=1}^{n}\frac{\hbar^{2}}{2m_{i}}\nabla_{i}\psi\cdot\nabla_{i}\psi^{*}+V\psi^{*}\psi,\label{eq:2.13}
\end{equation}
from which by the usual procedures one arrives at the $n$-particle
Schrödinger equation 
\begin{equation}
\i\hbar\frac{\partial\psi}{\partial t}=\left(-\sum_{i=1}^{n}\frac{\hbar^{2}}{2m_{i}}\nabla_{i}^{2}+V\right)\psi.\label{eq:2.14}
\end{equation}
Note also that from \eqref{eq:2.9} one obtains upon variation in
$P$ the modified Hamilton-Jacobi equation familiar from the de\,Broglie-Bohm
interpretation, i.e., 
\begin{equation}
\frac{\partial S}{\partial t}+\sum_{i=1}^{n}\frac{(\nabla_{i}S)^{2}}{2m_{i}}+V(\VEC x_{1},\VEC x_{2},\dots,\VEC x_{n},t)+U(\VEC x_{1},\VEC x_{2},\ldots,\VEC x_{n},t)=0,\label{eq:2.15}
\end{equation}
where $U$ is known as the ``quantum potential'' 
\begin{equation}
U(\VEC x_{1},\VEC x_{2},\ldots,\VEC x_{n},t)=\sum_{i=1}^{n}\frac{\hbar^{2}}{4m_{i}}\left[\frac{1}{2}\left(\frac{\nabla_{i}P}{P}\right)^{2}-\frac{\nabla_{i}^{2}P}{P}\right]=-\sum_{i=1}^{n}\frac{\hbar^{2}}{2m_{i}}\frac{\nabla_{i}^{2}R}{R}\;.\label{eq:2.16}
\end{equation}
Moreover, with the definitions 
\begin{equation}
\VEC u_{i}:=\frac{\delta\VEC p_{i}}{m_{i}}=-\frac{\hbar}{2m_{i}}\frac{\nabla_{i}P}{P}\quad\text{and}\quad\VEC{k_{u}}_{i}:=-\frac{1}{2}\frac{\nabla_{i}P}{P}=-\frac{\nabla_{i}R}{R}\;,\label{eq:2.17}
\end{equation}
one can rewrite $U$ as 
\begin{equation}
U=\sum_{i=1}^{n}\left[\frac{m_{i}\VEC u_{i}\cdot\VEC u_{i}}{2}-\frac{\hbar}{2}(\nabla_{i}\cdot\VEC u_{i})\right]=\sum_{i=1}^{n}\left[\frac{\hbar^{2}}{2m_{i}}(\VEC{k_{u}}_{i}\cdot\VEC{k_{u}}_{i}-\nabla_{i}\cdot\VEC{k_{u}}_{i})\right]\;.\label{eq:2.18}
\end{equation}
However, as was already detailed in Grössing (2009)~\cite{Groessing.2009origin},
for the energy balance $kT=\hbar\omega$ referring to the vacuum's
acting as the particle's ``thermostat'', $\VEC u_{i}$ can also
be written as 
\begin{equation}
\VEC u_{i}=\frac{1}{2\omega_{i}m_{i}}\nabla_{i}Q\;,\label{eq:2.19}
\end{equation}
which thus explicitly shows its dependence on the spatial behavior
of the heat flow $Q=kT\ln P$. Insertion of \eqref{eq:2.19} into
\eqref{eq:2.18} then provides a thermodynamic formulation of the
quantum potential as 
\begin{equation}
U=\sum_{i=1}^{n}\frac{\hbar^{2}}{4m_{i}}\left[\frac{1}{2}\left(\frac{\nabla_{i}Q}{\hbar\omega_{i}}\right)^{2}-\frac{\nabla_{i}^{2}Q}{\hbar\omega_{i}}\right]\;.\label{eq:2.20}
\end{equation}

\section{``Systemic'' Nonlocality in Double Slit Interference\label{sec:Dynamical-Nonlocality-in}}

In a recent paper, Tollaksen\emph{~et\,al}.\ \cite{Tollaksen.2010quantum}
renew the discussion on interference at a double slit ``from a single
particle perspective'', asking the following question: If a particle
goes through one slit, how does it ``know'' whether the second slit
is open or closed? We shall here first recapitulate the arguments
providing these authors' answer and later provide our own arguments
and answer. Of course, the question is about the phase information
and how it affects the particle. We know from quantum mechanics that
phases cannot be observed on a local basis and that a common overall
phase has no observational meaning. Assuming that two Gaussian wave
functions, $\Psi_{1}$ and $\Psi_{2}$, describe the probability amplitudes
for particles emerging from slits $1$ or $2$, respectively, which
are separated by a distance $D,$ the total wave function for the
particle exiting the double slit may be written as
\begin{equation}
\Psi=\e^{\i\alpha_{1}}\Psi_{1}+\e^{\i\alpha_{2}}\Psi_{2},\label{eq:dyn.3.0}
\end{equation}
but since a common overall phase is insignificant, one multiplies
\eqref{eq:dyn.3.0} with $\e^{-\i\alpha_{1}}$ and writes the total
wave function as
\begin{equation}
\Psi_{\varphi}=\Psi_{1}+\e^{\i\varphi}\Psi_{2},
\end{equation}
where $\varphi:=\alpha_{2}-\alpha_{1}$ is the physically significant
\emph{relative phase}. Tollaksen\emph{~et\,al}.\ now ask where
the relative phase appears in the form of (deterministic) observables
that describe interference. For the following recapitulation of their
argument, it must be stressed that the authors assume the two wave
functions, $\Psi_{1}$ and $\Psi_{2}$, to be initially non-overlapping.
When looking at the expectation value (in the one-dimensional case,
for simplicity)
\begin{align}
\overline{x}=\int\Psi_{\varphi}^{*}x\Psi_{\varphi}\d x & =\int\left(\Psi_{1}^{*}+\e^{-\i\varphi}\Psi_{2}^{*}\right)x\left(\Psi_{1}+\e^{\i\varphi}\Psi_{2}\right)\d x\\
 & =\int\left(\Psi_{1}^{*}x\Psi_{1}+\Psi_{2}^{*}x\Psi_{2}\right)\d x+\int\Psi_{1}^{*}x\e^{\i\varphi}\Psi_{2}\d x+\mathrm{c.c.},\nonumber 
\end{align}
one sees that it is independent of $\varphi$ because of the vanishing
of the last term due to the assumed non-overlapping of $\Psi_{1}$
and $\Psi_{2}$. Similarly, this also holds for all moments of $x,$
and for all moments of $p$ as well. In particular, one has for the
expectation value of the momentum
\begin{align}
\overline{p} & =\mathrm{Re}\int\Psi_{\varphi}^{*}\i\hbar\frac{\partial}{\partial x}\Psi_{\varphi}\d x\label{eq:dyn.3.1-0}\\
 & =\mathrm{Re}\int\left(\Psi_{1}^{*}\i\hbar\frac{\partial}{\partial x}\Psi_{1}+\Psi_{2}^{*}\i\hbar\frac{\partial}{\partial x}\Psi_{2}\right)\d x+\mathrm{Re}\int\Psi_{1}^{*}\i\hbar\frac{\partial}{\partial x}\e^{\i\varphi}\Psi_{2}\d x+\mathrm{c.c.},\nonumber 
\end{align}
where the $\varphi$--dependent term vanishes identically, because
$\frac{\partial}{\partial x}\Psi_{2}=0$ for $\Psi_{2}=0$. So, again,
where does the relative phase appear? The answer of Tollaksen\emph{~et\,al}.\ is
given by a ``shift operator'' that shifts the location of, say $\Psi_{1}$,
over the distance $D$ to its new location coinciding with that of
$\Psi_{2}$. The expectation value of the shift operator is thus given
by
\begin{align}
\overline{\e^{-\i\frac{pD}{\hbar}}}= & \int\left(\Psi_{1}^{*}\e^{-\i\frac{pD}{\hbar}}\Psi_{1}+\Psi_{2}^{*}\e^{-\i\frac{pD}{\hbar}}\Psi_{2}\right)\d x\\
 & +\int\Psi_{1}^{*}\e^{-\i\frac{pD}{\hbar}}\Psi_{2}\e^{\i\varphi}\d x+\int\Psi_{2}^{*}\e^{-\i\varphi}\e^{-\i\frac{pD}{\hbar}}\Psi_{1}\d x,\nonumber 
\end{align}
where all but the last term vanish identically, thus providing (with
the correct normalization)
\begin{equation}
\overline{\e^{-\i\frac{pD}{\hbar}}}=\e^{-\i\varphi}/2\;\textrm{ and }\;\overline{\e^{-\i\frac{pD}{\hbar}}}+\overline{\e^{\i\frac{pD}{\hbar}}}=\cos\varphi.
\end{equation}
In order to physically interpret the shift operator, the authors now
shift to the Heisenberg picture, thereby providing with a Hamiltonian
$H=\frac{p^{2}}{2m}+V\left(x\right)$ the time evolution of the operator
as
\begin{equation}
\frac{\d}{\d t}\:\e^{-\i\frac{pD}{\hbar}}=\frac{\i}{\hbar}\left[\e^{-\i\frac{pD}{\hbar}},\frac{p^{2}}{2m}+V\left(x\right)\right]=\frac{-\i D}{\hbar}\e^{-\i\frac{pD}{\hbar}}\left\{ \frac{V\left(x\right)-V\left(x+D\right)}{D}\right\} .\label{eq:dyn.3.1-1}
\end{equation}
With its dependence on the distance $D$ between the two slits, Eq.
\eqref{eq:dyn.3.1-1} is a description of \emph{dynamical nonlocality}:
it is thus shown how a particle can ``know'' about the presence
of the other slit. Tollaksen\emph{~et\,al}.\ maintain that it is
possible to understand this dynamical nonlocality only by employing
the Heisenberg picture. However, we shall now show that such an understanding
is possible also within the Schrödinger picture, and even more intuitively
accessible, too. For, there is one assumption in the foregoing analysis
that is not guaranteed to hold in general, i.e., the non-overlapping
of the wave functions $\Psi_{1}$ and $\Psi_{2}$. On the contrary,
we now shall assume that the two Gaussians representing the probability
amplitudes for the particle immediately after passing one of the two
slits do not have any artificial cut-off, but actually extend across
the whole slit system, even with only very small (and practically
often negligible) amplitudes in the regions further away from the
slit proper. (We shall give arguments for this assumption further
below.) In other words, we now ask: what if $\Psi_{1}$ and $\Psi_{2}$
do overlap, even if only by a very small amount? To answer this question,
we consider the expectation value of the momentum and obtain from
Eq. \eqref{eq:dyn.3.1-0} that the terms involving the relative phase
$\varphi$ provide with $\Psi_{j}=R_{j}e^{\i\varphi_{j}}$ and $\varphi_{j}=\frac{S_{j}}{\hbar}$
\begin{equation}
\overline{p}=R_{1}R_{2}\left(\nabla S_{1}+\nabla S_{2}\right)\cos\varphi+\hbar\left(R_{1}\nabla R_{2}-R_{2}\nabla R_{1}\right)\sin\varphi.\label{eq:dyn.3.1-2}
\end{equation}
First of all one notes upon comparison of \eqref{eq:dyn.3.1-2} with
Eqs.~\eqref{eq:3.12}--\eqref{eq:3.13} the exact correspondence
of $\overline{p}$ with our classically obtained expression for the
interference terms of the emerging current, or of the expression for
$m\boldsymbol{\mathbf{\overline{v}}}_{\mathrm{tot}}$, respectively.
Moreover, although the product $R_{1}R_{2}$ is in fact negligibly
small for regions where only a long tail of one Gaussian overlaps
with another Gaussian (i.e., such that the non-overlapping assumption
would be largely justified), nevertheless the second term in \eqref{eq:dyn.3.1-2}
can be very large despite the smallness of $R_{1}$ or $R_{2}$. It
is this latter part which is responsible for the genuinely ``quantum
mechanical'' nature of the average momentum, i.e., for its nonlocal
nature. This is formally similar in the Bohmian picture, but here
given a more direct physical meaning in that this last term refers
to a difference in diffusive currents as explicitly formulated in
the last term of Eq.~\eqref{eq:3.12}. Because of the mixing of diffusion
currents from both channels, we call this dominant term in $\mathbf{J_{\mathrm{tot}}}$
the ``entangling current''\cite{Mesa.2013variable}, $\mathbf{J_{\mathrm{e}}}$.
In fact, inserting Eq.~\eqref{eq:2.19} for a one-particle system
into the latter expression, one obtains
\begin{equation}
\mathbf{J_{\mathrm{e}}}=\sqrt{P_{1}P_{2}}\frac{\nabla\left(Q_{1}-Q_{2}\right)}{2\omega m}\sin\varphi_{12}.\label{eq:50}
\end{equation}
In other words, the entangling current is essentially a measure for
the gradient of the vacuum's ``heat'' fluxes $Q_{i}$ associated
with the channels $i$. 

Finally, there are substantial arguments against the non-overlapping
scenario in Tollaksen\emph{~et\,al}. Firstly, experiments by Rauch\emph{~et\,al}.\ have
shown that in interferometry interference does not only happen when
the main bulks of the Gaussians overlap, but also when a Gaussian
interferes with the off-bulk plane-wave components of the other wave
function as well \cite{Rauch.1993phase}. On a theoretical side, we
have repeatedly stressed that the diffusion processes employed in
our model must be described by nonlocal diffusion wave fields \cite{Mandelis.2001diffusion-wave,Mandelis.2001structure}
which thus require small but non-zero amplitudes across the whole
experimental setup. Moreover, and more specifically, we have shown
\cite{Groessing.2010emergence,Groessing.2011dice} that quantum propagation
can be identified with sub-quantum anomalous (i.e., ``ballistic'')
diffusion which is characterized by infinite mean displacements $\overline{x}=\infty$
despite the finite drifts $\overline{x^{2}}=u^{2}t^{2}$ and $u<c.$
In sum, these arguments speak in favor of using small, but non-zero
amplitudes from the Gaussian of the ``other slit'' which can interfere
with the Gaussian at the particle's location.

Let us now see how we can explain dynamical nonlocality within our
sub-quantum approach. For example, we can ask the following question
in our context: what if we start with one slit only, and when the
particle should pass it we open the second slit? Let us assume for
the time being, and without restriction of generality, that the $x$--component
of the velocity $v_{x}$ is zero. Then, according to \eqref{eq:dyn.2.19},
upon opening the second slit (with the same $\sigma$), we obtain
a term proportional to the distance $2X$ between the two slits. Thus,
there will be a shift in momentum on the particle passing the first
slit given by
\begin{equation}
\frac{\Delta p}{\hbar}=\pm\frac{1}{2}\nabla\varphi_{12}=\pm\frac{1}{2\hbar}\nabla\left(S_{1}-S_{2}\right)=\frac{\Delta p_{\mathrm{mod}}}{\hbar},\label{eq:dyn.4.1}
\end{equation}
where one effectively uses the ``modular momentum'' $p_{\mathrm{mod}}=p\textrm{ mod}\frac{h}{2X}=p-2n\pi\frac{\hbar}{2X}$
because an added or subtracted phase difference $\varphi_{12}=2n\pi$
does not change anything. In other words, by splitting $X$ into a
component $X_{n}$ providing $\varphi_{12}=2n\pi$ on the one hand,
and the modular remainder $\Delta X$ on the other hand providing
$\varphi_{12}=p_{\mathrm{mod}}x/\hbar$, one rewrites \eqref{eq:dyn.2.19}
with $X:=X_{n}+\Delta X$ (and with $v_{x}=0$ for simplicity) as
\begin{equation}
\varphi_{12}=-(X_{n}+\Delta X)x\frac{1}{D}\frac{u^{2}t}{\sigma_{0}^{2}}=:-2n\pi-\Delta Xx\frac{\hbar}{2m}\frac{t}{\sigma^{2}\sigma_{0}^{2}}\:.\label{eq:dyn.4.2}
\end{equation}

Therefore, one can further on substitute $\Delta p$ by $\Delta p_{\mathrm{mod}}$,
and one obtains a momentum shift (with the sign depending on whether
the right or left of the two slits is opened as the second one)
\begin{equation}
\Delta p_{\mathrm{mod}}=\pm\frac{\hbar}{2}\nabla\varphi_{12}=\pm\frac{\Delta X}{2}\frac{\hbar^{2}}{2m}\frac{t}{\sigma^{2}\sigma_{0}^{2}}=\pm m\Delta X\frac{D^{2}t}{\sigma^{2}\sigma_{0}^{2}}=\pm m\Delta X\frac{\dot{\sigma}}{\sigma}\:.\label{eq:dyn.4.3}
\end{equation}
It follows that, due to the ``vacuum pressure'' stemming from the
opened second slit, there is an \emph{emergent} nonlocal ``force''
which does not derive from a potential but from the impinging of the
second slit's sub-quantum diffusive momenta on the particle at the
first slit. As there is no additional force on the particle, we do
not speak of a ``dynamical'' nonlocality, but of a ``systemic''
one instead. For more details, see Grössing\emph{~et\,al}.\ (2013)
\cite{Groessing.2013dice}.

\section{Quantum Mechanical Results Obtained from Classical Computer Simulations
\label{sec:Computer Simulations}}

For our classical simulations of quantum phenomena we make use of
the physics of ballistic diffusion as given in Eq.~\eqref{eq:2.24},
with the time-dependent diffusivity $D_{{\rm t}}$ from Eq. \eqref{eq:2.25}.
We use an explicit finite difference forward scheme\cite{Schwarz.2009numerische},
\begin{align}
\frac{\partial P}{\partial t} & \rightarrow\frac{1}{\Delta t}\left(P[x,t+1]-P[x,t]\right),\\
\frac{\partial^{2}P}{\partial x^{2}} & \rightarrow\frac{1}{\Delta x^{2}}\left(P[x+1,t]-2P[x,t]+P[x-1,t]\right),
\end{align}
 with 1-dimensional cells. For our case of $D_{{\rm t}}$ being independent
of $x$, the complete equation after reordering reads as
\begin{equation}
P[x,t+1]=P[x,t]+\frac{D[t+1]\Delta t}{\Delta x^{2}}\left\{ P[x+1,t]-2P[x,t]+P[x-1,t]\right\} \label{eq:tdde.1.2.3}
\end{equation}
with space $x$ and time $t$, and initial Gaussian distribution $P(x,0)$
with standard deviation $\sigma_{0}$.

In the following, some exemplary figures with one spatial and one
time dimension are shown to demonstrate some of the present applications
of our model as well as of the novel simulation protocol that comes
along with it. The first three figures exhibit double-slit scenarios,
with time evolving from bottom to top, and the remaining three figures
show \emph{n}-slit scenarios with time running from left to right.
Except for the last figure, the coloring for intensity distributions
is, for increasing intensity, from white through yellow and orange,
and average trajectories are displayed in red.

\begin{figure}
\begin{centering}
\includegraphics[width=9cm]{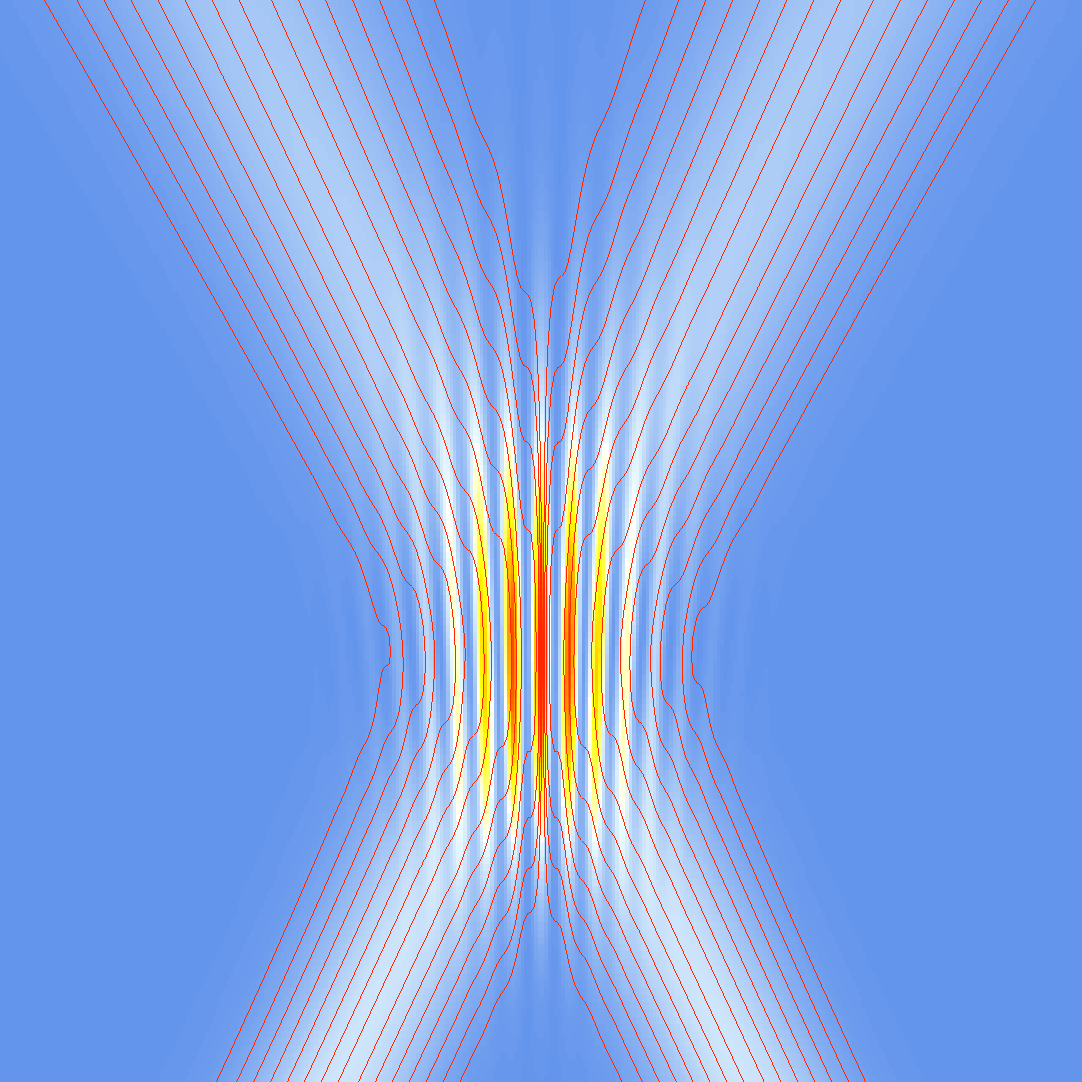}
\par\end{centering}

\vspace*{8pt}

\begin{centering}
\includegraphics[width=9cm]{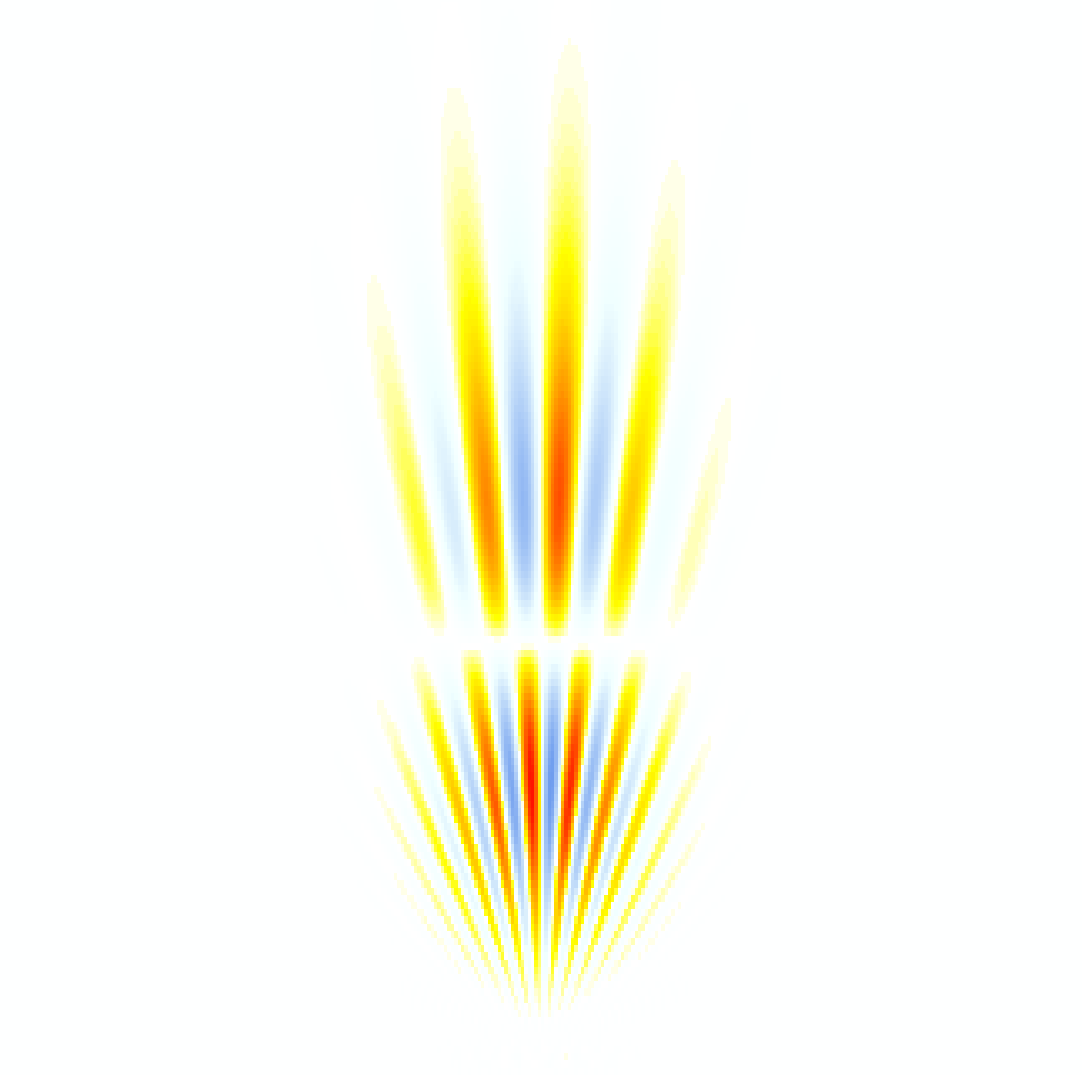}
\par\end{centering}

\vspace*{8pt}

\caption{\label{fig:2a}\emph{Top:} Quantum mechanical interference pattern,
with averaged trajectories, obtained from a computer simulation employing
classical physics only. \emph{Bottom:} ``Entangling current'' of
Eq. \eqref{eq:50} for the scenario above. }
\end{figure}

Turning now to Fig.\ref{fig:2a}, we begin with the upper part, i.e.,
a\emph{ }quantum mechanical interference pattern obtained from a computer
simulation employing classical physics only. Shown are the probability
density distributions for two wave packets emerging from the Gaussian
slits with opposite velocities, superimposed by flow lines which coincide
with Bohmian trajectories. Note, however, that the latter appear only
as the result of the averaging, whereas the sub-quantum model on which
the figure is based involves wiggly, stochastic motions. For the smoothed-out
trajectories, however, a ``no crossing'' rule applies with respect
to the central symmetry axis, just like in the Bohmian picture. The
figure's lower part shows the ``entangling current'' of Eq.~\eqref{eq:50}
for the scenario above, which is in our model the decisive part of
the probability density current that is responsible for the genuinely
``quantum'' effects. Due to the sinusoid nature, colors are used
that display positive (red) and negative (blue) values. Note that
after the time of maximal overlap between the two wave-packets, the
order of maxima and minima is reversed, which results from the respective
differences in the exchanges of ``heat'' according to Eq.~\eqref{eq:50}. 

\begin{figure}
\begin{centering}
\includegraphics[width=9cm]{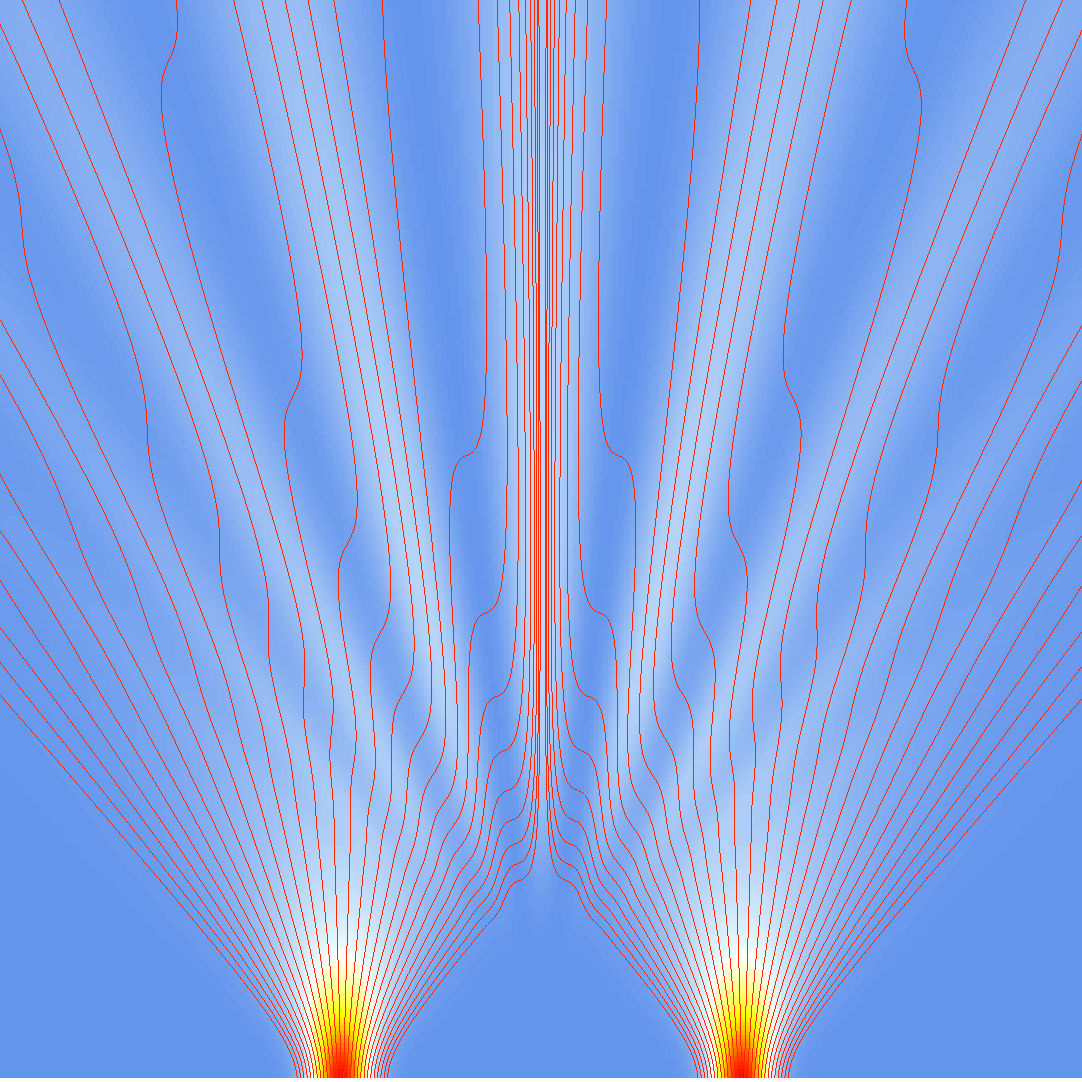}
\par\end{centering}

\vspace*{8pt}

\begin{centering}
\includegraphics[width=9cm]{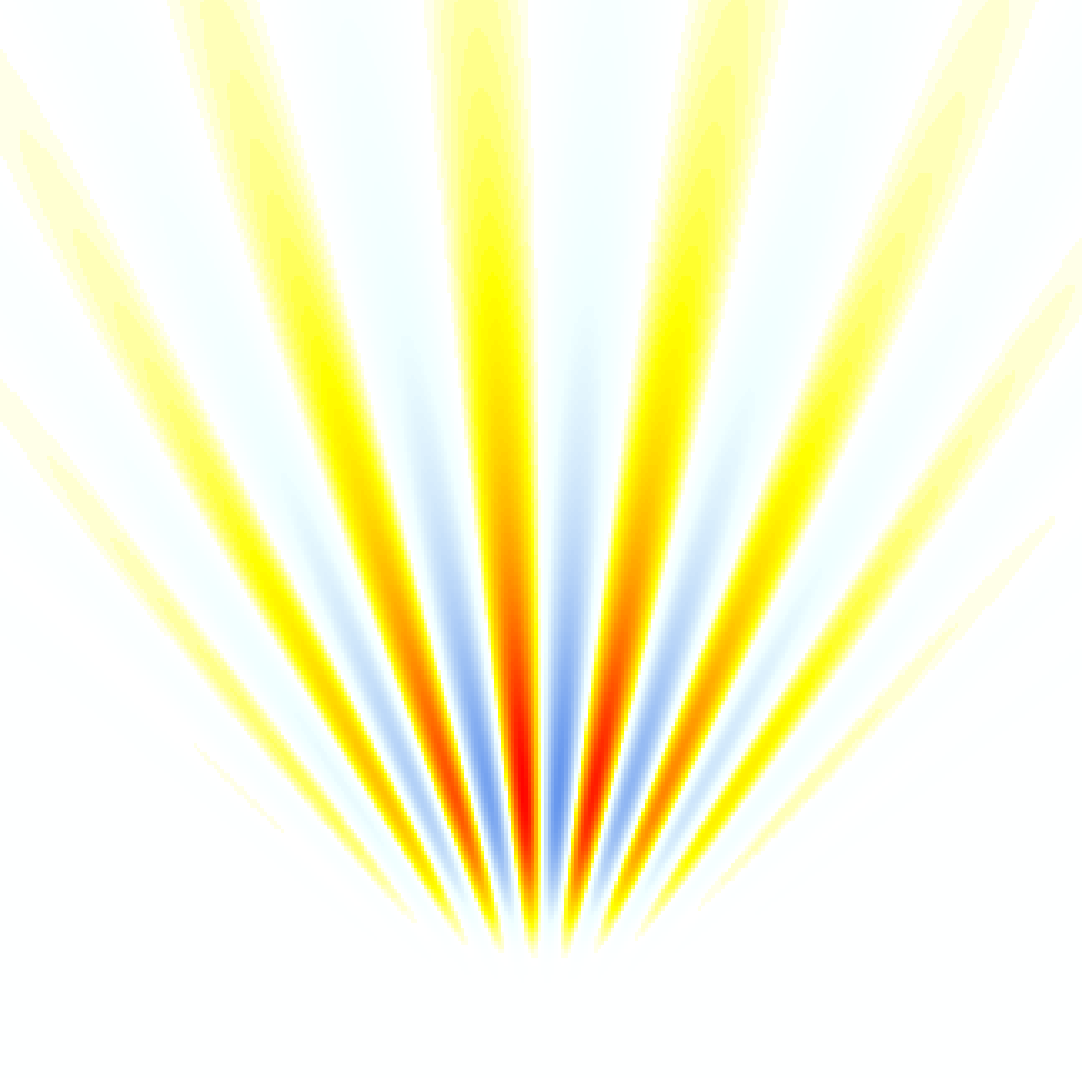}
\par\end{centering}

\vspace*{8pt}

\caption{\label{fig:fig3}Similar scenario as in Fig.\ref{fig:2a}, albeit
with zero velocity components in the transverse (i.e., $x-$) direction,
and with large dispersion instead of a small one. As can be seen from
the entangling current in the lower part of the figure, its extrema
coincide with the areas where the trajectories' ``kinks'' appear,
pointing at a rapid crossing-over due to the effects of the diffusive
processes involved.}
\end{figure}

Fig.\ref{fig:fig3} shows a similar scenario as Fig.\ref{fig:2a},
albeit with zero velocity components in the transverse (i.e., $x-$)
direction. However, whereas the situation of Fig.\ref{fig:2a} is
characterized by small dispersion of the Gaussians, here the dispersion
is significantly higher. The effects of the ``no crossing'' rule
are clearly visible. Note also the ``kinks'' of trajectories moving
from the center-oriented side of one relative maximum to cross over
to join more central (relative) maxima. In our model, a detailed micro-causal
account of the corresponding kinematics can be given. As can be seen
from the entangling current in the lower part of the figure, its extrema
coincide with the areas where the kinks appear, pointing at a rapid
crossing-over due to the effects of the diffusive processes involved.

\begin{figure}
\begin{centering}
\includegraphics[width=9cm]{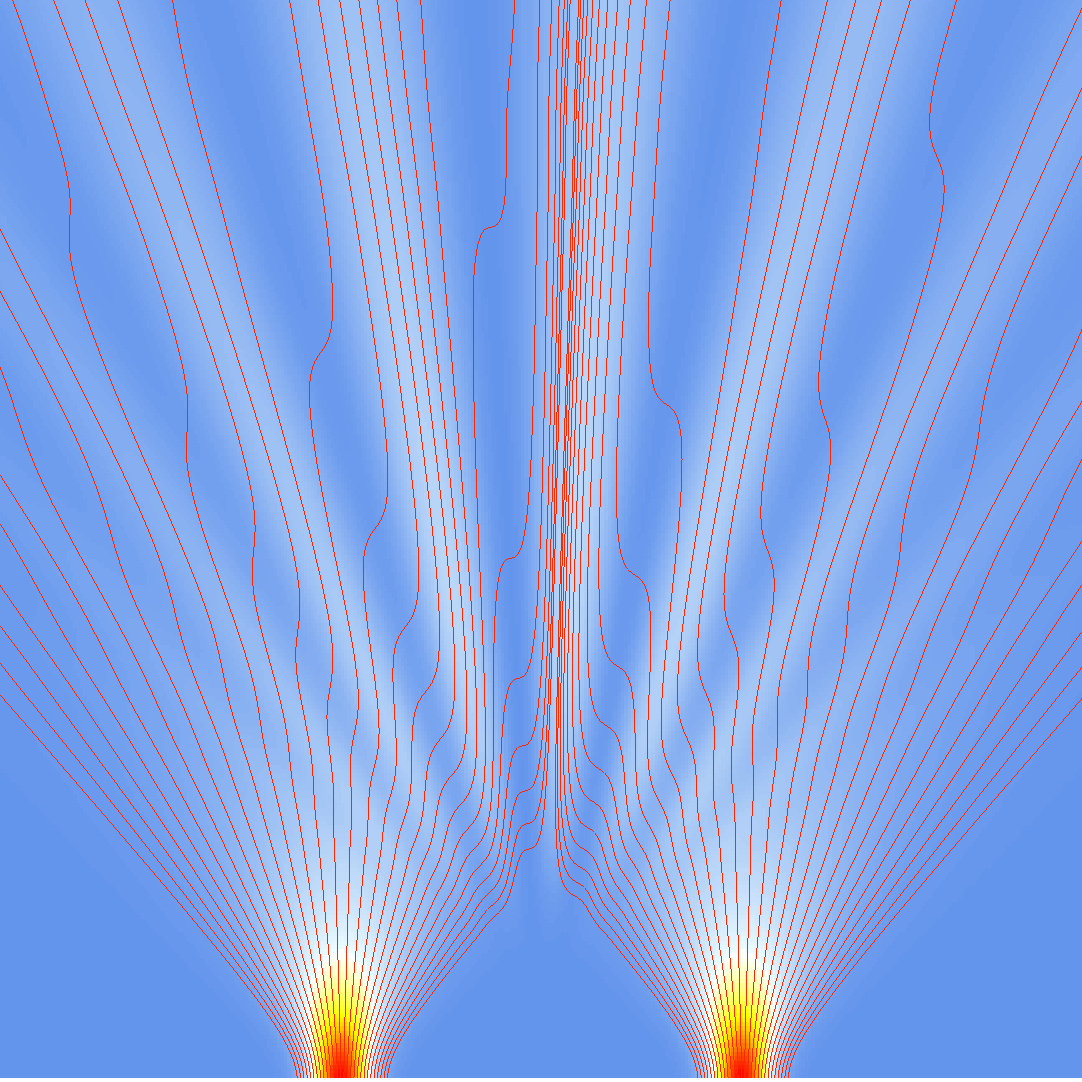}
\par\end{centering}

\vspace*{8pt}

\begin{centering}
\includegraphics[width=9cm]{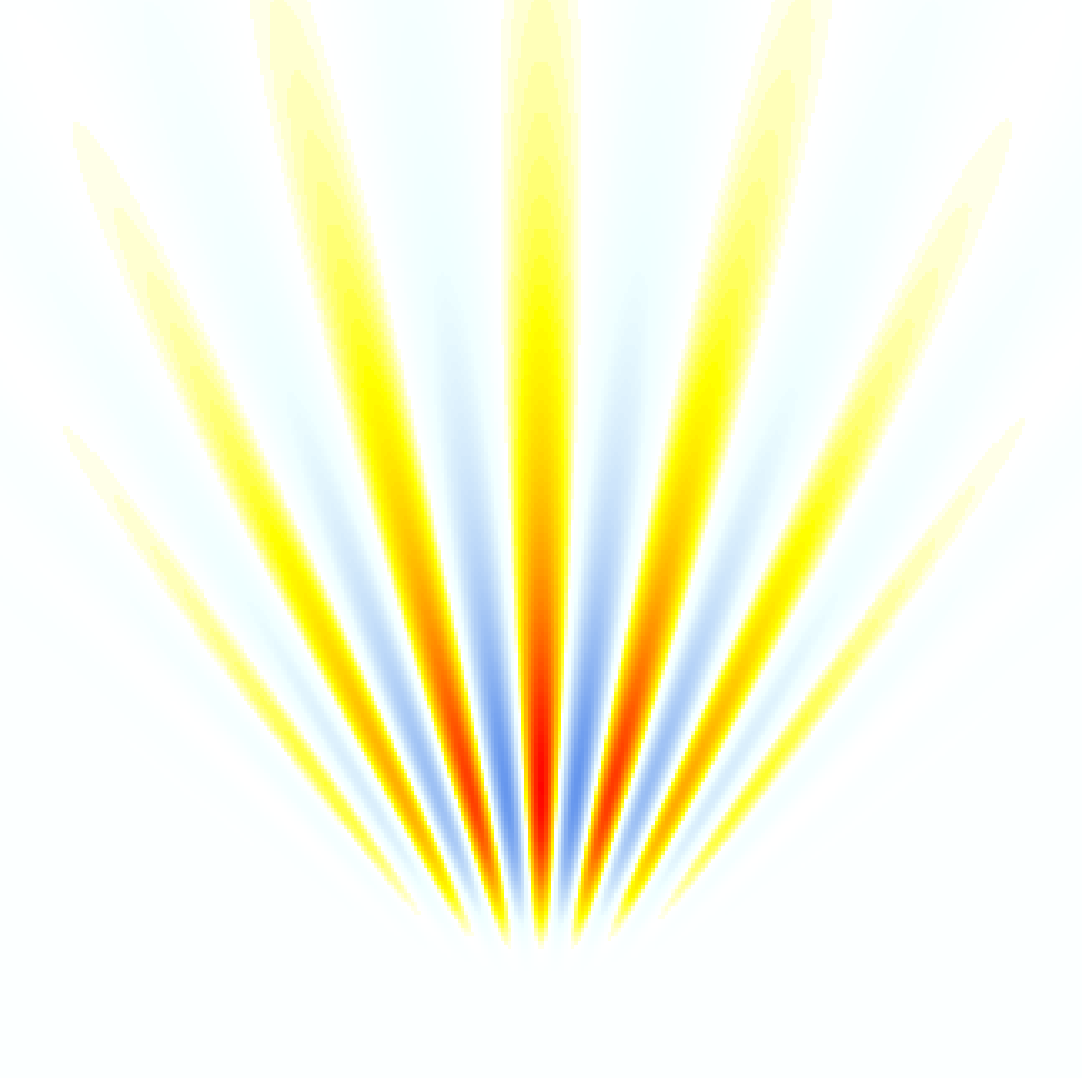}
\par\end{centering}

\vspace*{8pt}

\caption{\label{fig:delta-phi-pi-2}Same as Fig.\ref{fig:fig3}, except that
in the left channel a phase shift of $\triangle\varphi=\frac{\pi}{2}$
has been added, thus demonstrating an example of the scalar Aharonov-Bohm
effect. Due to the entangling current, or to the diffusion processes,
respectively, the intensity maxima of Fig.\ref{fig:fig3} are ``pushed
to the side'', providing the asymmetrical distribution of the interference
fringes. }
\end{figure}

Fig.\ref{fig:delta-phi-pi-2} shows almost the same scenario as Fig.\ref{fig:fig3},
the only difference being that in the left channel a phase shift of
$\triangle\varphi=\frac{\pi}{2}$ has been added, essentially leading
to a representation of the scalar Aharonov-Bohm effect. Note that
the minima and maxima of the intensity distribution are shifted accordingly.
This comes along with a different behavior of the entangling current.
Whereas in Fig.\ref{fig:fig3} the entangling current is asymmetrical
with respect to the central symmetry line, it now is symmetrical.
Effectively, this means that due to the diffusion processes providing
some ``vacuum pressure'', the intensity maxima are ``pushed to
the side'', providing an asymmetrical distribution of the interference
fringes. 

\begin{figure}
\begin{centering}
\includegraphics[width=9cm]{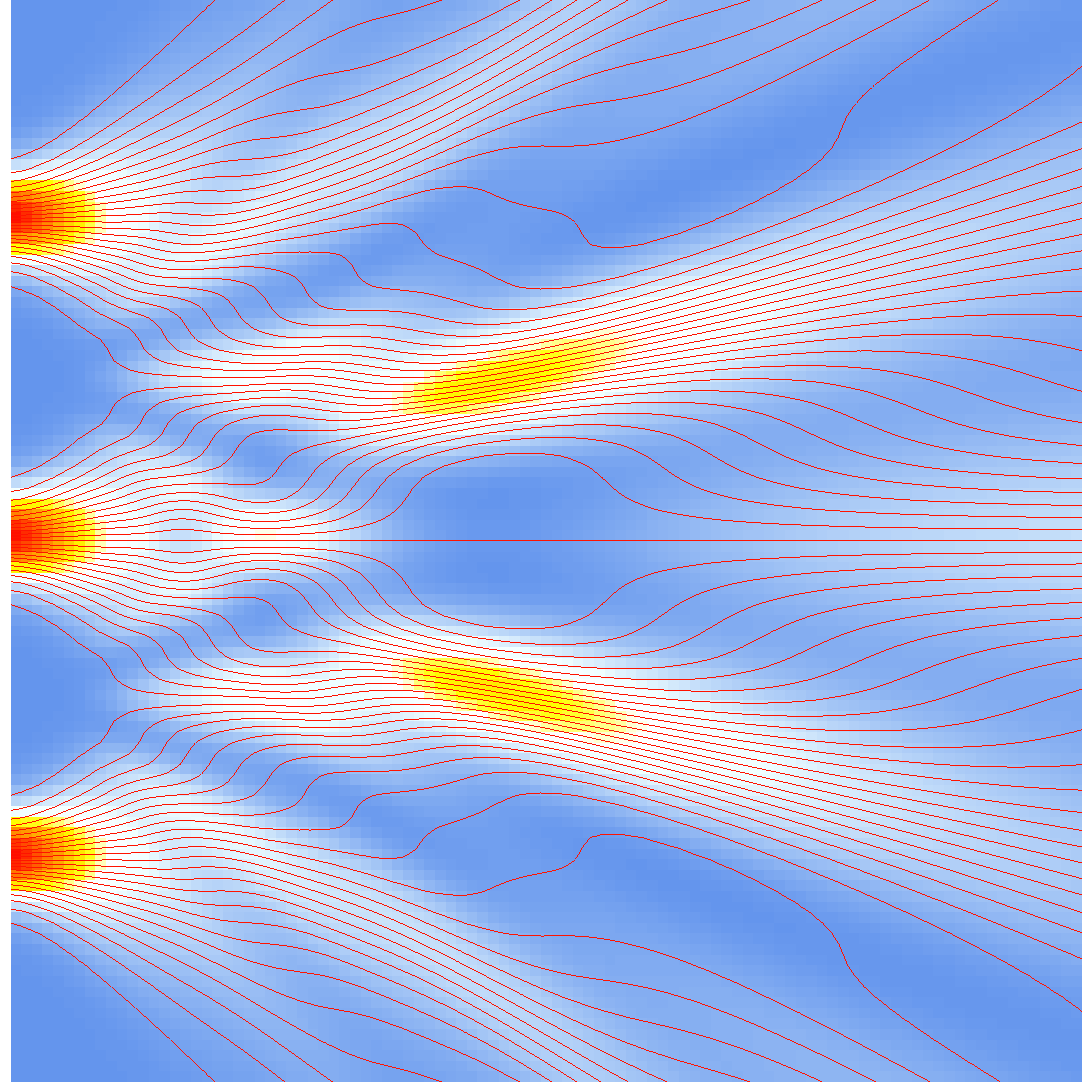}
\par\end{centering}

\vspace*{8pt}

\begin{centering}
\includegraphics[width=9cm]{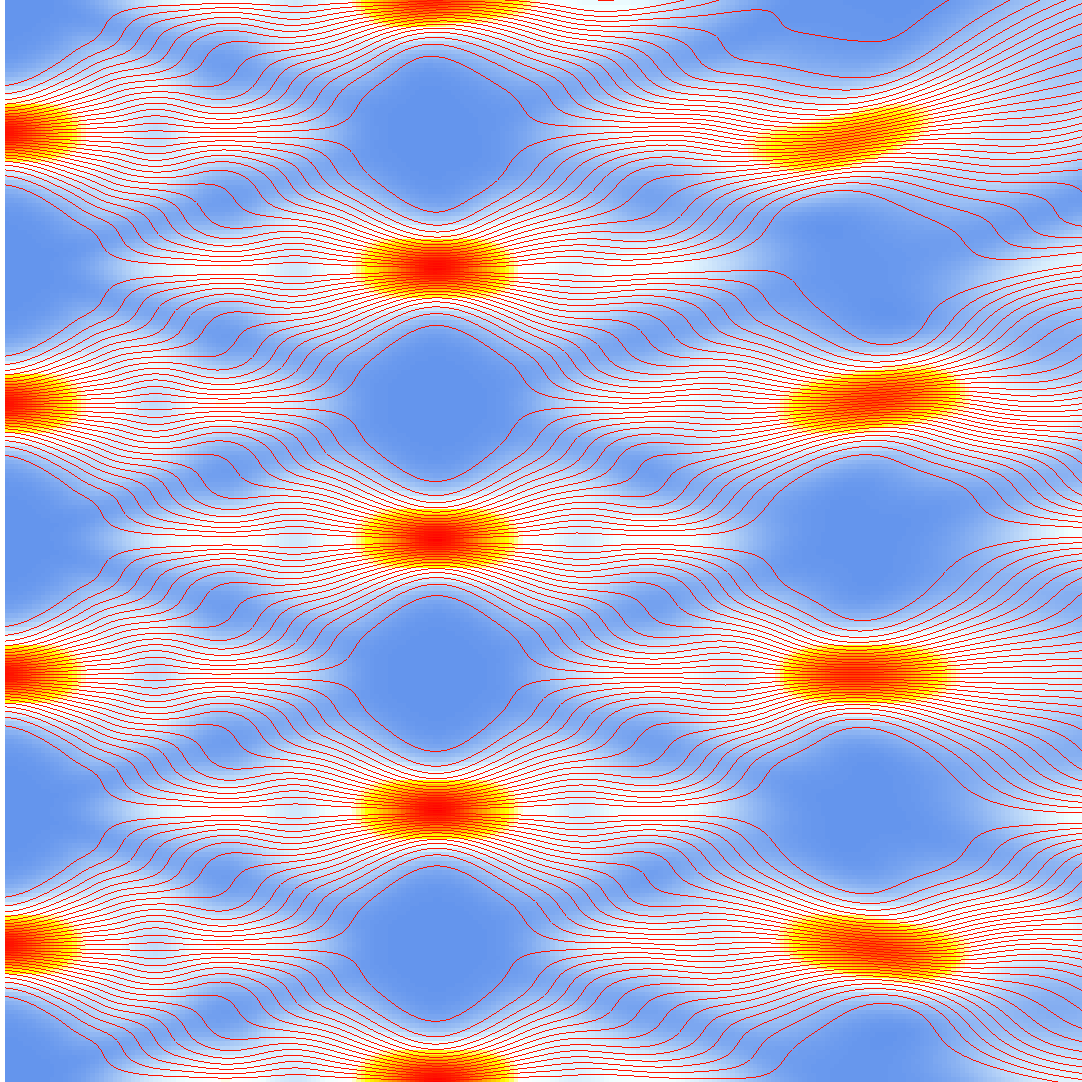}
\par\end{centering}

\vspace*{8pt}

\caption{\label{fig:test3a}\emph{Top:} A 3-slit system with intensity distributions
and averaged trajectories. \emph{Bottom:} Detail of a system with
a large number $n$ of slits (i.e., a so-called Talbot carpet), explicitly
showing only four of them. Note that non-spreading ``cells'' emerge
which contain the trajectories and keep them there.}
\end{figure}

Fig.\ref{fig:test3a} displays two examples which extend the double
slit simulations to more than two slits. Whereas the upper picture
shows a 3-slit system with intensity distributions and averaged trajectories,
the lower one is a detail of a ``Talbot carpet'', i.e., a system
with a large number $n$ of slits, albeit exhibiting only four of
them. It clearly demonstrates a result that is well-known from comparable
Bohmian calculations, i.e., that for the case of large $n$, non-spreading
``cells'' emerge which contain the trajectories and keep them there
\cite{Sanz.2007causal}.

\begin{figure}
\begin{centering}
\includegraphics[width=9cm]{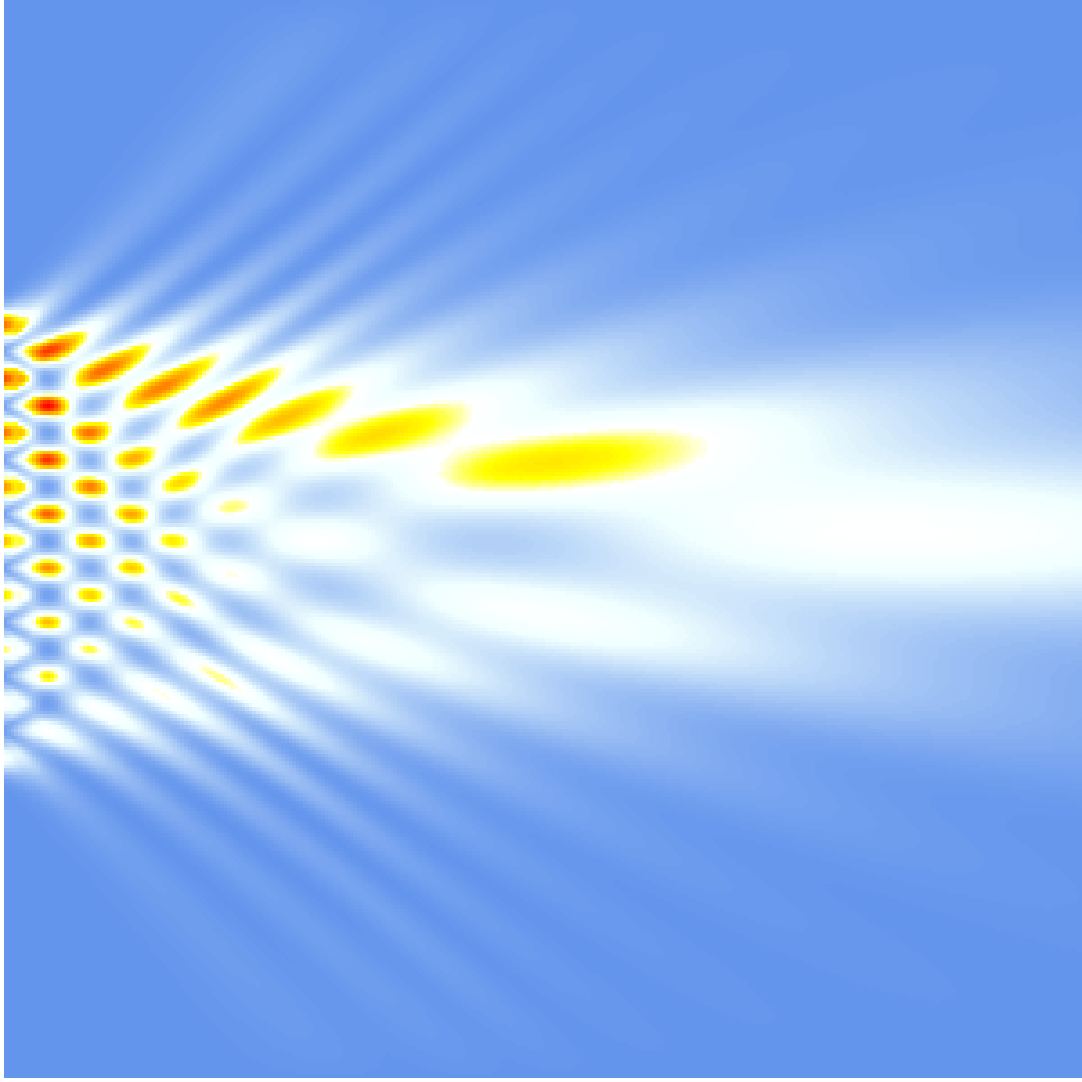}
\par\end{centering}

\vspace*{8pt}

\begin{centering}
\includegraphics[width=9cm]{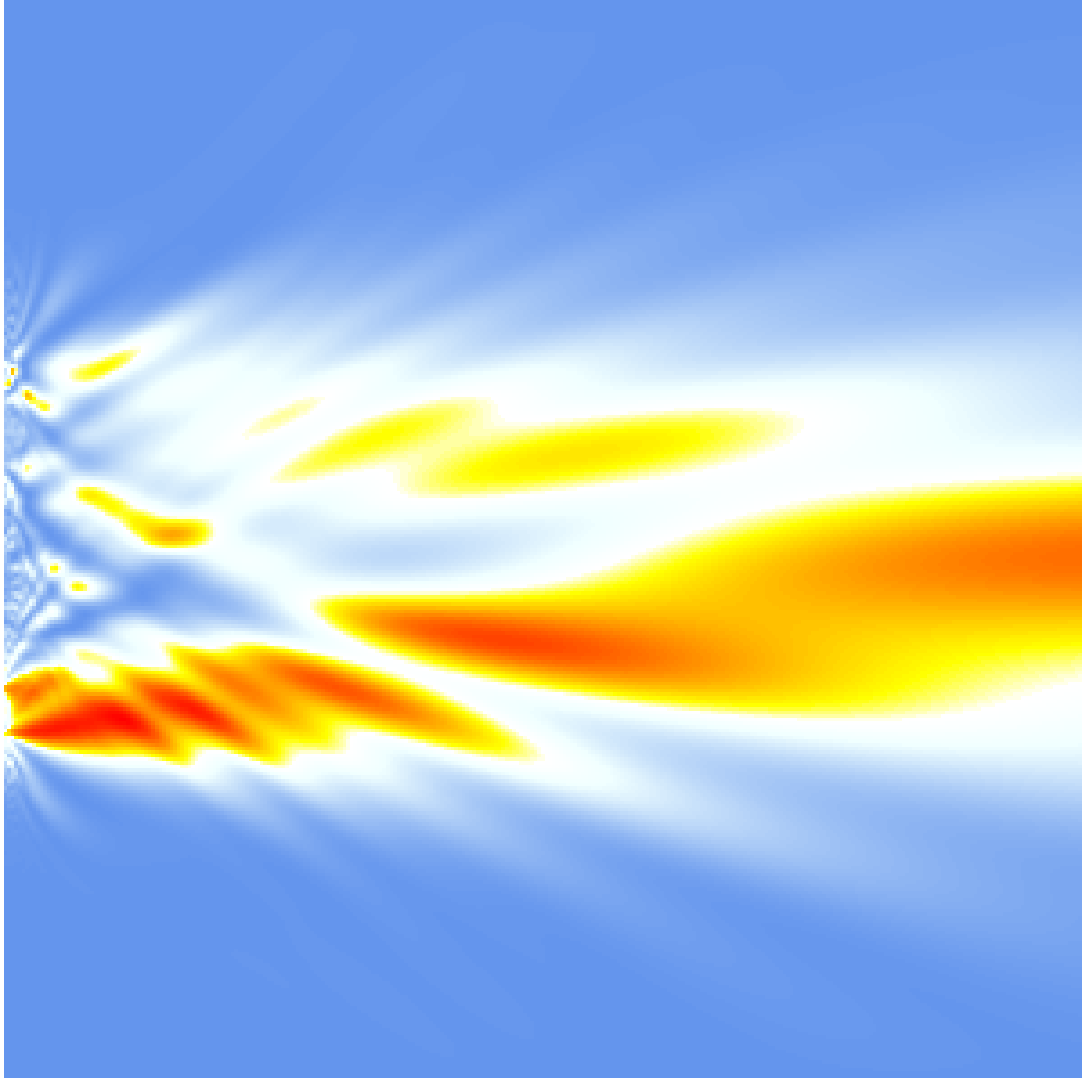}
\par\end{centering}

\vspace*{8pt}

\caption{\label{fig:vaxjo-9a}\emph{Top:} Classically simulated interference
pattern for a system of 9 slits with gradually diminishing initial
intensity distributions. \emph{Bottom:} Same, except that the initial
intensity distributions have been chosen to have random values.}
\end{figure}

To continue with $n$-slit systems, Fig.\ref{fig:vaxjo-9a} shows,
in the upper part, a type of Talbot carpet (without trajectories)
with gradually diminishing initial intensity distributions, from top
to bottom. The lower figure shows the same except that the initial
intensity distributions have been chosen to have random values. This
is an example of a system whose complexity is considerable, despite
the fact that the simulation is still very fast and simple, thus illustrating
also an advantage over purely analytical tools. Eventually, this type
of simulations might even apply at levels of complexity which would
exceed any traditional analytical methods of quantum mechanics.

\begin{figure}
\begin{centering}
\includegraphics[height=9cm]{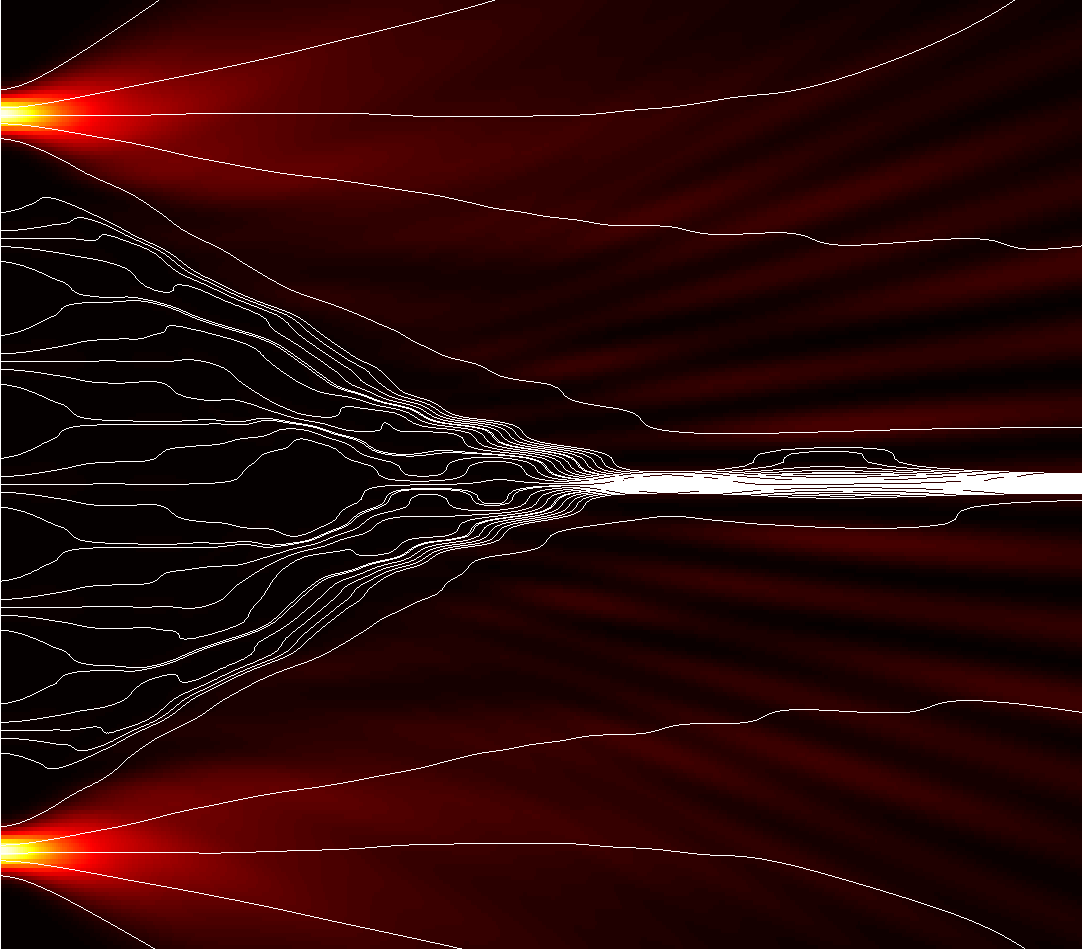}
\par\end{centering}

\vspace*{8pt}

\begin{centering}
\includegraphics[height=9cm]{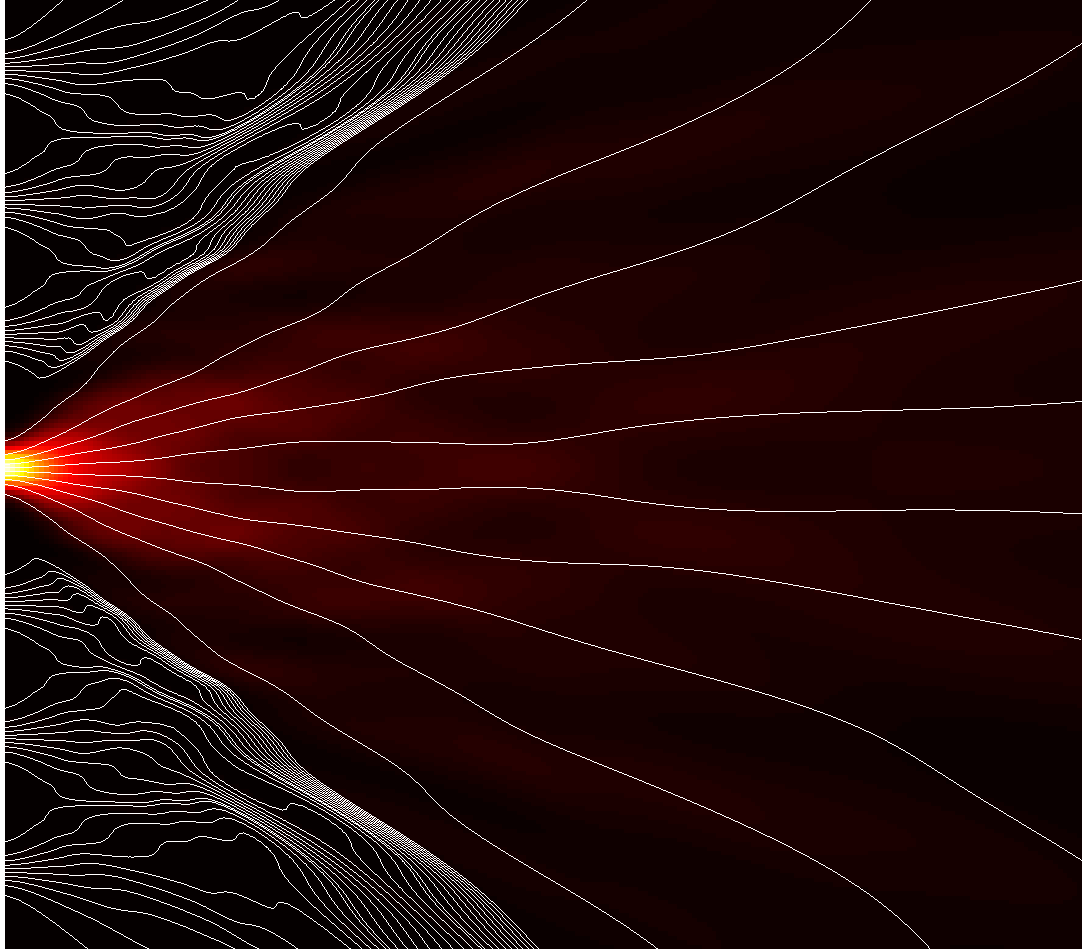}
\par\end{centering}

\vspace*{8pt}

\caption{\label{fig:qupressure1a}Two examples of intensity and trajectory
distributions of particularly ``weighted'' 7-slit systems. \emph{Top:}
Quantum squeezer. \emph{Bottom:} Quantum sweeper.}
\end{figure}

Finally, Fig.\ref{fig:qupressure1a} shows two examples of intensity
and trajectory distributions of particularly ``weighted'' 7-slit
systems. In the upper figure, the high relative intensities at the
extreme locations force the trajectories of the remaining slits to
become ``squeezed'' into a narrow canal. The lower figure instead
shows the opposite effect of a centrally positioned relatively high
intensity ``sweeping'' away the other slits' trajectories. The two
examples again illustrate the effect of the vacuum's ``pressure''
in our sub-quantum model.

\section*{Acknowledgments}

I want to thank my friends and AINS colleagues Siegfried Fussy, Johannes
Mesa Pascasio, and Herbert Schwabl for their continued enthusiasm
and support in trying to work out a radically new approach to quantum
mechanics. Further, I thank the organizers of the summer school, and
particularly Claudio Furtado, Inacio Pedrosa, Claudia Pombo, and Theo
Nieuwenhuizen for inviting me to Joao Pessoa, for their perfect organization
and their great hospitality. Finally, I also gratefully acknowledge
partial support by the Fetzer Franklin Fund.

\providecommand{\href}[2]{#2}\begingroup\raggedright\endgroup


\end{document}